\documentstyle[12pt]{article}

\newcommand{\secn}[1]{Section~\ref{#1}}
\newcommand{\bra}[1]{\langle{#1}|}
\newcommand{\ket}[1]{|{#1}\rangle}
\newcommand{\braket}[2]{\langle{#1}|{#2}\rangle}
\newcommand{\tbl}[1]{Table~\ref{#1}}
\newcommand{\eq}[1]{Eq.~(\ref{#1})}
\newcommand{\fig}[1]{Fig.~\ref{#1}}
\newcommand{\nl}{\nonumber \\}
\def\zeroslash{\mathord{\not\mathrel{\hskip 0.04 cm 0}}}
\def\ex{{\rm e}}
\def\beq{\begin{equation}}
\def\eeq{\end{equation}}
\def\beqa{\begin{eqnarray}}
\def\eeqa{\end{eqnarray}}
\newcommand{\sect}[1]{\setcounter{equation}{0}\section{#1}}
\renewcommand{\theequation}{\thesection.\arabic{equation}}
\newcommand{\EQ}{\begin{equation}}
\newcommand{\EN}{\end{equation}}
\newcommand{\bea}{\begin{eqnarray}}
\newcommand{\ena}{\end{eqnarray}}
\newcommand{\vs}[1]{\vspace{#1 mm}}
\renewcommand{\a}{\alpha}
\renewcommand{\b}{\beta}
\renewcommand{\c}{\gamma}
\renewcommand{\d}{\delta}
\newcommand{\e}{\epsilon}
\newcommand{\ve}{\varepsilon}
\newcommand{\hs}[1]{\hspace{#1 mm}}
\newcommand{\shalf}{\frac{1}{2}}
\renewcommand{\Im}{{\rm Im}\,}
\newcommand{\NP}[1]{Nucl.\ Phys.\ {\bf #1}}
\newcommand{\PL}[1]{Phys.\ Lett.\ {\bf #1}}
\newcommand{\NC}[1]{Nuovo Cimento {\bf #1}}
\newcommand{\CMP}[1]{Comm.\ Math.\ Phys.\ {\bf #1}}
\newcommand{\PR}[1]{Phys.\ Rev.\ {\bf #1}}
\newcommand{\PRL}[1]{Phys.\ Rev.\ Lett.\ {\bf #1}}
\newcommand{\MPL}[1]{Mod.\ Phys.\ Lett.\ {\bf #1}}
\newcommand{\IJMP}[1]{Int.\ Jour.\ of\ Mod.\ Phys.\ {\bf #1}}
\newcommand{\dpb}{D$p$-brane}
\newcommand{\dpbs}{D$p$-branes}
\renewcommand{\thefootnote}{\fnsymbol{footnote}}
\def\one{{\hbox{ 1\kern-.8mm l}}}

\def\R{{\rm R}}
\def\ii{{\rm i}}

\newlength{\bredde}
\def\slash#1{\settowidth{\bredde}{$#1$}\ifmmode\,\raisebox{.15ex}{/}
\hspace*{-\bredde} #1\else$\,\raisebox{.15ex}{/}\hspace*{-\bredde} #1$\fi}

\renewcommand{\baselinestretch}{1.0}
\renewcommand{\thefootnote}{\arabic{footnote}}

\begin{document}
%\begin{titlepage}
%\rightline{DFTT 6/98}
\rightline{NORDITA 1999/84-HE}
%\rightline{KUL-TF-98/10}
%\rightline{hep-th/9912275}
%\leftline{very preliminary draft}
\rightline{\hfill December 1999}
\vskip 1.2cm
%\vskip 0.8cm
\centerline{\Large \bf D branes in string theories, II \footnote{Work 
partially supported by the European Commission
TMR programme ERBFMRX-CT96-0045 and by 'Programma di breve mobilit{\`{a}}
per scambi internazionali dell'Universit{\`{a}} di Napoli "Federico II" '.}}
\vskip 0.8cm
\centerline{\bf P. Di Vecchia$^a$
%\footnote{e-mail: divecchia@nbivms.nbi.dk} 
and A. Liccardo$^b$}
\vskip .5cm
\centerline{\sl $^a$ NORDITA, Blegdamsvej 17, DK-2100 Copenhagen \O, Denmark}
%\vskip .2cm
%\centerline{\sl $^b$ Dipartimento di Fisica Teorica, Universit\`a di
%Torino} 
%\centerline{\sl and I.N.F.N., Sezione di Torino, Via P. Giuria 1, I-10125 
%Torino, Italy}
%\vskip .2cm
%\centerline{\sl $^c$ Dipartimento di Scienze e Tecnologie Avanzate}
%\centerline{\sl Universit\`a del Piemonte Orientale, I-15100 
%Alessandria, Italy} 
\vskip .2cm
\centerline{\sl $^b$ Dipartimento di Fisica, Universit\`a di Napoli 
and I.N.F.N.,}
\centerline{\sl  Sezione di Napoli, 
Mostra d'Oltremare Pad. 19, I-80125 Napoli, Italy}
\vskip 1.3cm
\centerline{Lectures presented at the YITP Workshop on ``Developments in }
\centerline{ Superstring and M-theory'', Kyoto, Japan , October 1999.} 
\begin{abstract}
In these lectures we review the properties of a boosted and rotated 
boundary state and of a boundary state with an abelian gauge field
deriving from it the Dirac-Born-Infeld action and  a newly
constructed class  of classical  solutions. We also review the construction
of the boundary state for the stable non-BPS state of type I theory
corresponding to the perturbative state 
present at the first excited level  of the $SO(32)$ heterotic string  and 
transforming according to the spinor representation of $SO(32)$.
\end{abstract}
%\end{titlepage}
%\newpage

\sect{Introduction}
\label{intro}

D$p$-branes are classical solutions of the eqs. of motion of the low-energy
string effective action~\footnote{See Ref.~\cite{REV} and references therein.}, 
charged under a $(p+1)$-form R-R field, that 
correspond to new non-perturbative BPS states of string theory, break $1/2$ 
supersymmetry, and are required by T-duality in theories with open 
strings~\footnote{See Ref.~\cite{lectPOL} and references therein.}.
They are characterized by the fact that open strings have
their end-points attached to them~\cite{POL95} and are conveniently described 
by a state of closed string theory called the boundary state. A review of
their properties and of the origin and of the construction of the 
boundary state can be found in Ref.~\cite{island}. For the sake of
completeness we rewrite here its explicit form. 
It is given by
\beq
|B, \eta \rangle _{R,NS} = \frac{T_p}{2} |B_{mat}, \eta \rangle  | B_{g},
\eta \rangle~~~~;~~~~T_p = \sqrt{\pi} \left( 2 \pi \sqrt{\alpha'} \right)^{3-p}
\label{bounda3}
\eeq
where the boundary states for the matter part and for the ghost degrees of
freedom are given by
\beq
|B_{mat} \rangle  = |B_X \rangle  |B_{\psi}, \eta \rangle
\hspace{1cm};\hspace{1cm}
|B_{g}\rangle  = |B_{gh} \rangle  | B_{sgh}, \eta \rangle
\label{bounda4}
\eeq
The ghost part of the boundary state can be found in eqs.(6.216) and 
(7.251)-(7.253) of Ref.~\cite{island}.
Here we will only write the explicit form of the matter part
of the boundary state. The part corresponding  to the bosonic coordinate $X$ 
is equal to
\beq
\label{b1}
|B_X \rangle  = \delta^{d-p-1}({\hat q}^i-y^i) \left(\prod_{n=1}^\infty
e^{-\frac{1}{n}
\alpha_{-n}
S\cdot\widetilde\alpha_{-n}}\right)|0\rangle _{\alpha}|0\rangle
_{\widetilde\alpha}
|p=0\rangle ~,
\eeq
where
\beq
\label{matS}
S^{\mu \nu} = ( \eta^{\alpha \beta} , - \delta^{ij} )
\eeq
$\alpha, \beta$ are indices along the world volume of the D$p$-brane, while 
$i,j$ span the transverse directions of the brane.

The fermionic part of the matter boundary state is equal to
\beq
\label{bnsns}
|B_{\psi} , \eta \rangle  = -i \prod_{t=1/2}^\infty 
\left(e^{i\eta\psi_{-t}\cdot S\cdot
\widetilde \psi_{-t}} \right) |0\rangle
\eeq
in the NS-NS sector and to
\beq
\label{brr}
|B_{\psi} , \eta \rangle  =-\prod_{t=1}^\infty e^{i\eta\psi_{-t}\cdot
S\cdot\widetilde
\psi_{-t}} |B_{\psi} , \eta \rangle ^{(0)}
\eeq
in the R-R sector. The zero mode contribution $|B_{\psi}, \eta 
\rangle ^{(0)}$ is given by
\beq
\label{solover}
|B_\psi , \eta\rangle ^{(0)} = {\cal M}_{AB}|A\rangle |\widetilde B\rangle
\eeq
where
\beq
\label{soloverM}
{\cal M}_{AB}=\left(C\Gamma^0...\Gamma^p\frac{1+i\eta\Gamma^{11}}{1+i\eta}
\right)_{AB}
\eeq
$C$ is the charge conjugation matrix and $\Gamma^\mu$ are the Dirac
$\Gamma$ matrices in the $10$-dimensional space. Their properties are 
summarized in the Appendix. The boundary state in 
eq.(\ref{bounda3}) depends on the two values of $\eta = \pm 1$. Actually we
must take a combination of them  corresponding to the
GSO projection. The GSO-projected states are given by:
\beq
|B \rangle_{NS} = \frac{1}{2} \left( |B, + \rangle_{NS} - |B, -\rangle_{NS} 
\right)~~~;~~~
|B\rangle_{R} = \frac{1}{2} \left( |B, +\rangle_{R} + |B, -\rangle_{R} \right) 
\label{gso}
\eeq
respectively for the NS and R sectors.

In Ref.~\cite{island}  we have reviewed the properties of the D$p$-branes, 
described by the boundary state discussed above, and considered  as static 
and rigid objects  to which open strings are attached. We have not discussed 
the fact that they can be boosted, rotated
and that the excitations of the attached open strings provide dynamical
degrees of freedom to them. In particular the massless excitations that have
the property of not
changing the energy of the brane, can be interpreted as collective 
coordinates of the D$p$-branes. In these lectures we fill this gap by
showing how to construct a boosted~\cite{CANGEMI}~\footnote{See also 
Ref.~\cite{KLEBA}.} and 
rotated boundary state and a 
boundary state 
containing a constant abelian gauge field living in the world volume of
the brane~\cite{CALLAN}. We then show that some of those boundary states are 
related by 
T-duality. We then use the boundary state with an external gauge field in
order, on the one hand, to derive the Born-Infeld action and, on the other 
hand, to reconstruct, with a specific choice of the external field, newly found
solutions~\cite{LU1,LU2,LU3} of the eqs. of motion of the low-energy string 
effective action. Finally we  discuss  the boundary state for a toroidally 
compactified
space-time and we use it for describing the properties of stable non BPS 
states recently discussed in the literature~\footnote{For reviews on stable
non BPS states see Ref.~\cite{REVSEN}.}.

These lectures, that are partially based on the Ph.D. thesis of Antonella
Liccardo, are organized as follows. In sect.~\ref{boostrot} we discuss the 
properties
of the boosted and rotated boundary state. Sect.~\ref{extfield} is devoted 
to the boundary
state with an abelian gauge field. In sect.~\ref{traexci} we show how some 
of the previously
discussed boundary states are related by T-duality and we discuss the boundary
state with transverse excitations. In sect.~\ref{fdp} we show that, by 
choosing a
particular form of the gauge field, we reproduce the $(F,Dp )$ bound states.
Sect.~\ref{compact} is devoted to the construction of the boundary state in
the case of a compactified space-time and sect.~\ref{stable} to the 
construction of the boundary state for the stable
non-BPS particle of type I string theory. Finally in the Appendix we 
summarize the properties of the ten-dimensional $\Gamma$ matrices.

\vskip 1.2cm
\section {Boosted and Rotated Boundary State}
\label{boostrot}
In the introduction we have considered the boundary state corresponding
to a static {\dpb}. In this section we want to extend this construction  to
a boosted and a rotated one. It is easy to see that a boost along a 
longitudinal  direction of a {\dpb} does not modify the boundary state 
explicitly written in the introduction as expected from the Poincar{\'{e}}
invariance of the classical solution along the longitudinal directions of
the brane. We can therefore concentrate us on a 
boost
along one of the transverse directions and let us call this direction $k.$
The way to construct a boosted boundary state  is, as we have done in 
Ref.~\cite{island}, to start from the boundary conditions for an open string 
attached to such a
{\dpb} and  then translate them into the language of the closed string
channel through a conformal transformation and a
conformal rescaling (see Ref.~\cite{island} for details).

The boundary conditions for an open string attached to a {\dpb} boosted
with  velocity $v$ in the direction $k$  are~\cite{bachas}
\beq
\partial_{\sigma} X^{\alpha}|_{\sigma=0} =0 \hspace{2cm} \alpha=1, ...., p
\label{neu2}
\eeq
\beq
\label{neu3}
\partial_\sigma\left.\left(X^0+v X^k\right)\right|_
{\sigma=0}=0
\eeq
\beq
X^{i} |_{\sigma=0} = y^i   \hspace{2cm} i= p+1, ....D-1,~~~~{\rm
and}~~~i\neq k
\label{dir2}
\eeq
\beq
\label{dir4}
\left.\left(\frac {X^k +v X^0}{\sqrt {1-v^2}}\right)\right|_
{\sigma=0}=\frac{y^k}{\sqrt {1-v^2}}
\eeq
where $\vec y$ is a vector
belonging to the space transverse to the Dp-brane and therefore has zero
component along the time and the other world volume
directions of the Dp-brane. 
In the closed channel the previous conditions translate into the following
equations that characterize the boosted boundary state $|B,v,y\rangle $ 
in the bosonic string:
\beq
\partial_{\tau} X^{\alpha}|_{\tau=0}|B,v,y\rangle  =0 \hspace{2cm} \alpha=1,
...., p
\label{neuc2}
\eeq
\beq
\label{neuc3}
\partial_\tau\left.\left(X^0+v X^k\right)\right|_
{\tau=0}|B,v,y\rangle =0
\eeq
\beq
\left.\left(X^{i} - y^i\right) \right|_{\tau=0} |B,v,y\rangle  =0
\hspace{2cm} i= p+1, ....D-1,~~~~{\rm and}~~~i\neq k
\label{dirc2}
\eeq
\beq
\label{dirc4}
\left.\left(\frac {(X^k-y^k) +v X^0}{\sqrt {1-v^2}}\right)\right|_
{\tau=0}|B,v,y\rangle =0
\eeq
Instead of the last equation we can also have a less restrictive one
\beq
\label{dirc4bis}
\partial_\sigma\left.\left(X^k+v X^0\right)\right|_
{\tau=0}|B,v\rangle =0
\eeq
which corresponds to the case of a brane which is delocalized in the
$k$ direction.

The only  overlap conditions that differ from those of the static case given
in Ref.~\cite{island}  are
those in the directions of the boost, namely the time and the $k$ directions
and they are equal to
\beq
\label{boo1}
\left({\hat p}^0+v {\hat p}^k\right)|B,v,y\rangle =0
\eeq
\beq
\label{boo2}
\left[\left(\alpha^0_n+\widetilde\alpha^0_{-n}\right)
+v\left(\alpha^k_n+\widetilde\alpha^k_{-n}\right)  \right]|B,v,y\rangle =0
~~~~~\forall n\neq 0
\eeq
\beq
\label{boo3}
\frac{ {\hat q}^k + v {\hat q}^0 }{ \sqrt{1 - v^2}}|B,v,y\rangle =
\frac{y^k}{\sqrt{1-v^2}} |B,v,y\rangle
\eeq
\beq
\label{boo4}
\left[\left(\alpha^k_n-\widetilde\alpha^k_{-n}\right)
+v\left(\alpha^0_n-\widetilde\alpha^0_{-n}\right)  \right]|B,v,y\rangle =0
~~~~~\forall n\neq 0
\eeq
Let us now determine the explicit expression of the state  $|B,v,y\rangle $
which
fulfills all the previous conditions. For the zero mode part 
eq.(\ref{boo3}) tells us that the boundary state must contain a $\delta$
function of the type
\beq
\label{boo5}
\delta\left(\frac{q^k+v q^0-y^k}{\sqrt {1-v^2}}\right)=
\sqrt{1 - v^2}
\delta(q^k+ v q^0-y^k),
\eeq
Since the operator that acts on the boundary state in eq.(\ref{boo1})
commutes with the $\delta$-function in eq.(\ref{boo5}), in order to satisfy
eq.(\ref{boo1}), it is sufficient to write the zero mode part as follows:
\beq
\label{boo7}
\sqrt {1-v^2}\delta\left(q^k+v q^0-y^k\right) | p=0 \rangle
\eeq
It is easy to check that it satisfies both zero mode eqs.(\ref{boo1})
and (\ref{boo3}). Let us consider now the non zero modes part of the overlap
conditions. It is easy to see that, in order to satisfy eqs.(\ref{boo2}) and
(\ref{boo4}), the non-zero mode part of the boundary state must have the
following structure
\beq
\prod_{n=1}^{\infty} \left( e^{- \frac{1}{n}\alpha_{-n}\cdot M(v)\cdot
\widetilde\alpha_{-n}} \right) |0\rangle _\alpha |0\rangle _{\widetilde\alpha}
\label{nzero}
\eeq
where the matrix $M$ is obtained from the matrix $S$ in eq.(\ref{matS}) 
by substituting its elements
$(S_{00},S_{0k},S_{k0},S_{kk})$  with the correspondent ones
\beq
 {\it M}_{00} ={\it M}_{kk}=- \frac{1+v^2}{1 - v^2}\,\,\,\,\,;\,\,\,\,\,
{\it M}_{0k} ={\it M}_{k0}=-\frac{2v}{1 - v^2}
\label{m0j}
\eeq
Putting eqs.(\ref{boo7}) and (\ref{nzero}) together we get the final
expression for the boosted boundary state:
\[
|B,v,y\rangle =\frac{T_p}{2}\prod_{i=p+1,i\neq j}^{d-1}
\left[ 
\delta({\hat q}^i-y^i) \right] \sqrt{1 - v^2}\,\,
\delta(q^k+v~q^0-y^k) 
\]
\beq
\label{boo9}
e^{-\sum_{n=1}
^\infty\frac{1}{n}\alpha_{-n}\cdot M(v)\cdot\widetilde\alpha_{-n}}|0\rangle _\alpha
|0\rangle _{\widetilde\alpha} |p=0\rangle ~.
\eeq
We have fixed the normalization factor to be 
$T_p/2$ as in the static case, but the overlap conditions alone do not allow
to fix it uniquely. In general the boundary state in eq.(\ref{boo9}) could 
include an arbitrary function $N(v)$ of the physical velocity $v$ that can
only be determined by requiring agreement between the calculation of the 
interaction between two D-branes in the closed and open string channel.  
In this case, however, we have an independent way of uniquely fixing its
normalization by  applying to the static boundary state  $| B_X \rangle $ in
eq.(\ref{b1}) an operator that performs a boost 
along the direction $k$ transverse to the world volume of the D-brane
\beq
  |B, y , w\rangle  = e^{i w^k J^{0}{}_k} \;
  |B, {}^{( w)} y\rangle ~, 
  \label{boosted}
\eeq
where  $w$ is related to the physical velocity $v$ through the relation
\beq
\label{vw}
v={\rm tgh}w
\eeq
${}^{(w)}y^k = y^k {\rm cosh}w$ is the boosted position of the D-brane and
the generator of the Lorentz transformation is equal to
\beq
  J^{\mu \nu} = q^{\mu} p^{\nu} - q^{\nu} p^{\mu} -i
  \sum_{n=1}^{\infty} 
  \left( a_{n}^{\dagger \mu } a_{n}^{ \nu } - a_{n}^{\dagger \nu } 
    a_{n}^{ \mu } +  {\widetilde{a}}_{n}^{\dagger\mu}
{\widetilde{a}}_{n}^{ \nu} -
    {\widetilde{a}}_{n}^{\dagger \nu} {\widetilde{a}}_{n}^{ \mu}\right)~.
  \label{lorgen}
\eeq
with $a_n \sqrt{n} = \alpha_n$ and $a^{\dagger}_{n} \sqrt{n} = \alpha_{-n}$
with $n >0$.
After some algebra it can be seen that the boosted boundary state in
eq.(\ref{boosted}) can be written in the following form
\[
|B, y, w(v)\rangle =\frac{T_p}{2}\prod_{i=p+1,i\neq k}^{d-1}
\left[ 
\delta({\hat q}^i-y^i) \right] \frac{1}{{\rm cosh}w}
\delta(q^k+{\rm tgh}w~q^0-y^k) 
\]
\beq
\label{boo99}
e^{-\sum_{n=1}
^\infty\frac{1}{n}\alpha_{-n}\cdot M(w)\cdot\widetilde\alpha_{-n}}|0\rangle _\alpha
|0\rangle _{\widetilde\alpha} |p=0\rangle ~,
\eeq
that exactly coincides with the one given in eq.(\ref{boo9}), 
as it can be easily seen by making use of eq.(\ref{vw}). In this way we have
shown that the overall normalization of the boosted boundary state in 
eq.(\ref{boo9}) is correct.

The previous construction can be easily generalized to describe
a rotated {\dpb}.
Obviously the configuration of a {\dpb} embedded in a $d$-dimensional
space-time is invariant under rotations in the longitudinal 
space as well as in the transverse space. This implies that 
the boundary state is invariant under rotations in the planes
$ (\alpha,\beta)$ or $(i,j)$ $\forall \alpha,\beta\in\{ 0,...,p\}$
and $\forall i,j\in\{ p+1,...,d-1\}.$ This means that, in order to get a
new boundary state, we must consider a {\dpb} which is rotated 
with an angle $\omega$ in one of the planes specified by the directions 
$(\alpha,k)$.

The open string attached to this brane at $\sigma=0$ satisfies the boundary 
conditions
\beq
\partial_{\sigma} X^{\beta}|_{\sigma=0} =0 \hspace{2cm} 
\forall\beta\in\{0, ...., p\}~~{\rm and}~~\beta\neq\alpha 
\label{rot1}
\eeq 
\beq
\label{rot2}
\partial_\sigma\left.\left(X^\alpha\cos\omega+
 X^k\sin\omega\right)\right|_
{\sigma=0}=0
\eeq
\beq
X^{i} |_{\sigma=0} = y^i   \hspace{2cm} i= p+1, ....D-1,~~~~{\rm and}~~~i\neq k
\label{rot3}
\eeq
\beq
\label{rot4}
\left.\left(X^k\cos\omega
 - X^{\alpha} \sin\omega-y^k\cos\omega\right)\right|_
{\sigma=0}=0
\eeq
Then the overlap conditions that the rotated boundary
state must satisfy in the directions different from $(\alpha,k)$
are the same as in the unrotated case. On the other hand along
the directions $(\alpha,k)$ we must impose the following conditions:
\beq
\label{rot22}
\partial_\tau\left.\left(X^\alpha\cos\omega+
 X^k\sin\omega\right)\right|_
{\tau=0} | B, \omega, y \rangle =0
\eeq
and
\beq
\label{rot42}
\left.\left(X^k\cos\omega
 - X^{\alpha} \sin\omega-y^k\cos\omega\right)\right|_
{\tau=0} | B, \omega, y \rangle =0
\eeq
that in terms of the oscillators become:
\beq
\label{root1}
\left({\hat p}^\alpha \cos \omega+{\hat p}^k \sin
\omega\right)|B,\omega,y\rangle =0
\eeq
\beq
\label{root2}
\left[\left(\alpha^\alpha_n+\widetilde\alpha^\alpha_{-n}\right)
+\tan \omega~\left(\alpha^k_n+\widetilde\alpha^k_{-n}\right)
\right]|B,\omega,y\rangle =0
~~~~~\forall n\neq 0
\eeq
\beq
\label{root3}
\left({\hat q}^k \cos\omega- {\hat q}^\alpha\sin\omega\right)
|B,\omega,y\rangle =y^k\cos\omega|B,\omega,y\rangle
\eeq
\beq
\label{root4}
\left[\left(\alpha^k_n-\widetilde\alpha^k_{-n}\right)
-\tan \omega\left(\alpha^\alpha_n-\widetilde\alpha^\alpha_{-n}\right)
\right]
|B,\omega,y\rangle =0
~~~~~\forall n\neq 0
\eeq
Proceeding as in the previous case it is easy to see that the rotated
boundary
state has the following form:
\[
|B,\omega,y\rangle =\frac{T_p}{2} \prod_{i=p+1,i\neq k}^{d-1}
\left[ 
\delta({\hat q}^i-y^i) \right]
\delta(\cos \omega~q^k-\sin \omega~q^\alpha-  y^k\cos\omega) 
\]
\beq
\label{root6}
e^{-\sum_{n=1}
^\infty\frac{1}{n}\alpha_{-n}\cdot M(\omega)\cdot\widetilde\alpha_{-n}}|0
\rangle _\alpha
|0\rangle _{\widetilde\alpha}|p=0\rangle 
\eeq
where in this case the matrix $M$ is obtained from the matrix 
$S$ appearing in eq.(\ref{matS}) by substituting its elements 
$(S_{\alpha \alpha},S_{\alpha k},S_{k \alpha},S_{kk})$  with the 
correspondent elements 
\beq
\label{root5}
M_{(\alpha \alpha)}=-M_{(kk)}= 
\cos 2\omega\,\,\,\,\,;\,\,\,\,\,   
M_{(0k)} =M_{(k0)}=\sin 2\omega 
\eeq 
The previous boundary state for a rotated {\dpb} can again also be obtained
by acting on the boundary state given in eq.(\ref{b1}) with the rotation 
operator
\begin{equation}
  |B, y , \omega\rangle = e
^{i \omega^k J^{\alpha}{}_k} \;
  |B, {}^{( \omega)} y \rangle~,
  \label{rotot}
\end{equation}
where ${}^{(\omega)}y^k =y^k \cos\omega$
is the rotated position of the D-brane and
$J^{\mu \nu}$ is defined in eq.(\ref{lorgen}).
The previous considerations can be easily extended to the fermionic 
coordinate obtaining the boosted boundary state in the case of superstring.
We will not present here its detailed derivation as in the bosonic case, but
we write only its final form.  We get
\beq
\label{bnsnsvel}
|B_{\psi} , \eta \rangle  = -i \prod_{t=1/2}^\infty 
\left(e^{i\eta\psi_{-t}\cdot M \cdot
\widetilde \psi_{-t}} \right) |0\rangle
\eeq
in the NS-NS sector and 
\beq
\label{brrvel}
|B_{\psi} , \eta \rangle  =-\prod_{t=1}^\infty e^{i\eta\psi_{-t}\cdot
M \cdot\widetilde
\psi_{-t}} |B_{\psi} , \eta \rangle ^{(0)}
\eeq
in the R-R sector. The matrix $M$ is obtained from $S$ as in eq.(\ref{m0j}).
The zero mode contribution $|B_{\psi}, \eta \rangle ^{(0)}$ is given by
\beq
\label{solovervel}
|B_\psi , \eta\rangle ^{(0)} = \frac{1}{\sqrt{1-v^2}} 
\left(C [\Gamma^0 + v \Gamma^{k}] \Gamma^1 ...\Gamma^p
\frac{1+i\eta\Gamma^{11}}{1+i\eta} \right)_{AB}
|A\rangle |\widetilde B\rangle
\eeq
At the end of this section we write the interaction between two branes
moving with velocity $v$ relative to each other originally performed in 
Ref.~\cite{bachas} in the open string channel and then reproduced in 
Ref.~\cite{CANGEMI}
in the closed string channel. The total contribution of
the NS-NS sector is given by:
\[
A_{NS} = V_{NN-1} ( 8 \pi^2 \alpha')^{-NN/2}\,iv \int_{0}^{\infty}
dt \left( \frac{1}{t}\right)^{DD/2} \int_{-\infty}^{\infty} d \tau 
{ \rm e}^{ -( y^2 + v^2 \tau^2)/(2 \pi \alpha' t)} \times
\]
\beq
\times \left[\frac{\Theta_{3} ( \theta | i t) }{\Theta_{1} 
( \theta | i t)} \left(\frac{f_3}{f_1} \right)^{6 - \nu} \left(\frac{f_4}{f_2}
 \right)^{\nu} - \frac{\Theta_{4} ( \theta | i t) }{\Theta_{1} 
( \theta | i t)} \left(\frac{f_4}{f_1} \right)^{6 - \nu} \left(\frac{f_3}{f_2}
 \right)^{\nu}  \right]
\label{nsboo}
\eeq
where $\Theta_i ( \theta| it )$ is the Jacobi $\Theta$-function and its argument
is related to the velocity by:
\beq
\theta = \frac{1}{2 \pi i} \log \frac{1-v}{1+v}
\label{theta52}
\eeq
$\nu$ is equal to the number of mixed N-D boundary conditions for the open
strings with their endpoints on the two branes and the functions $f_i$ can
be found in eqs.(9.282) and (9.283) of Ref.~\cite{island}.
For the sake of simplicity we give the R-R contribution in the case of equal
branes. In this case we get
\[
A_{R} =  V_{p} ( 8 \pi^2 \alpha')^{-(p+1)/2}\,(-iv) \int_{0}^{\infty}
dt \left( \frac{1}{t}\right)^{(9-p)/2} \int_{-\infty}^{\infty} d \tau 
{ \rm e}^{ -( y^2 + v^2 \tau^2)/(2 \pi \alpha' t)} \times
\]
\beq
\frac{\Theta_{2} ( \theta | i t) }{\Theta_{1} ( \theta | i t)} 
\left(\frac{f_2}{f_1} \right)^{6} 
\label{rcon76}
\eeq
Summing the two contribution we get the following expression for equal D
branes:
\[
A = V_{p} ( 8 \pi^2 \alpha')^{-(p+1)/2}\,(-iv) \int_{0}^{\infty}
dt \left( \frac{1}{t}\right)^{(9-p)/2} \int_{-\infty}^{\infty} d \tau 
{ \rm e}^{ -( y^2 + v^2 \tau^2)/(2 \pi \alpha' t)} \times
\]
\beq
\times \left\{ \frac{\Theta_{3} ( \theta | i t) }{\Theta_{1} ( \theta | i t)} 
\left(\frac{f_3}{f_1} \right)^{6} - \frac{\Theta_{4} ( \theta | i t) }{
\Theta_{1} ( \theta | i t)} \left(\frac{f_4}{f_1} \right)^{6} -
\frac{\Theta_{2} ( \theta | i t) }{\Theta_{1} ( \theta | i t)} 
\left(\frac{f_2}{f_1} \right)^{6}\right\}
\label{velo98}
\eeq
Using known identities between $\Theta$-functions one can rewrite the previous
expression as:
\[
A = V_{p} ( 8 \pi^2 \alpha')^{-(p+1)/2}\,(-iv) \int_{0}^{\infty}
dt \left( \frac{1}{t}\right)^{(9-p)/2} \int_{-\infty}^{\infty} d \tau 
{ \rm e}^{ -( y^2 + v^2 \tau^2)/(2 \pi \alpha' t)} \times
\]
\beq
\times \frac{\left[ \Theta_1 ( \theta/2 | it )\right]^4}{
\Theta_1 ( \theta | it ) f_{1}^{9}}
\label{inte65}
\eeq
that is equal to the expression obtained in Ref.~\cite{GRGUT} using the 
light-cone boundary state. In the limit of small velocity we get
\beq
A \rightarrow V_{p+1} (2 \pi \alpha ')^{3 -p} \frac{\Gamma (7/2 - p/2 )}{ 
 (4 \pi)^{(p+1)/2}}  \frac{v^4}{y^{7-p}}
\label{smavel}
\eeq
that agrees with the calculation using M(atrix) theory performed in 
Ref.~\cite{DKPS} in the case of a D particle ($p=0$).

\vskip 1.2cm
\section {Boundary state with an external field}
\label{extfield}

Until now  we have considered the branes as static or
rigidly moving objects, a sort of geometrical hyperplanes having open
strings attached with their endpoints. We have completely neglected the 
dynamics of the open strings attached to the D branes. In this section we
will consider those excitations. We are in particular
interested to the zero mode (massless) excitations that do not change the
energy of the brane and that correspond to its collective coordinates.
In absence of Chan Paton factors the massless excitations of an open
string are described by a $U(1)$ gauge field.
In presence of a {\dpb} the ten-dimensional gauge vector field  splits into
a longitudinal vector field living in the world volume of the brane and a
transverse part correponding to  $d-p-1$ scalar fields that have the physical 
interpretation of coordinates of the brane. 
In this section we will neglect the scalar fields corresponding to the 
transverse components of the gauge field and concentrate on the construction of
the boundary state describing a D$p$-brane with an abelian gauge field living
on its world volume. Then we will use it for deriving the 
Dirac-Born-Infeld (DBI) action with the inclusion of the Wess-Zumino term, 
that is the action describing the low-energy dynamics of a D$p$-brane. 

In order to construct the boundary state with an abelian field on it  we 
follow the same procedure that we have followed in the previous section. 
We start
looking at the boundary conditions of an open string stretching between two
{\dpbs} with a gauge field on it and then  we translate these conditions
into those for the boundary state with a gauge field on it. 

Let us start considering the bosonic string.
An open string interacts with a gauge field through the pointlike charges
located at its endpoints. Therefore the action for an open string
interacting with an
abelian gauge field contains besides the free string action also a term
describing this interaction that occurs only at the  endpoints $\sigma=0$
and $\sigma=\pi$:
\beq
S= \int  d\tau \int_0^\pi d\sigma \left\{ \frac{1}{4\pi\alpha'}
[({\partial_{\tau} X})^2-({\partial_\sigma X})^2]-
[\delta(\sigma)-\delta(\sigma-\pi)]{\dot X}^\mu A_\mu(X)\right\}~.
\label{opact}
\eeq
For the sake of simplicity we have taken the two charges located at the 
endpoints of the string to have the same absolute value and opposite sign.
The eqs. of motion and the boundary conditions are obtained by varying the
previous action. One gets:
\[
 \delta S=-\frac{1}{2\pi\alpha'}\int  d\tau \int_0^\pi d\sigma\{
({\partial^2_\tau X_\mu}-{\partial^2_\sigma X_\mu})\delta X^\mu+
\partial_\sigma(\partial_\sigma X_\mu\delta X^\mu)+
\]
\beq
\label{varis}
+2\pi\alpha'
[\delta(\sigma)-\delta(\sigma-\pi)]
(-\delta X^\mu \partial_\tau A_\mu(X)+\partial_\tau X^\mu \delta A_\mu(X))\}
\eeq
Requiring this variation to be zero we obtain, together with the
equations of motion that are of course unchanged with respect to the free
case,
also the following boundary constraint
\beq
\label{sfinbou}
\frac{1}{2\pi\alpha'}\int d\tau \left\{
\partial_\sigma X_\mu  -
%[\delta(\sigma)-\delta(\sigma-\pi)]
\partial_\tau X^\rho {\hat{F}}_{\mu \rho} \right\}\delta X^\mu 
|^{\sigma =\pi}_{\sigma=0} =0
\eeq
where we have defined ${\hat A}^\mu=2\pi\alpha' A^\mu$ and ${\hat{F}}_{\mu \nu}
= \partial_{\mu} {\hat{A}}_{\nu} - \partial_{\nu} {\hat{A}}_{\mu}$.
We will now assume that the gauge field is non zero only in the directions
of the world volume of the brane, while it is constant along the transverse 
ones: $A^i= const$. This means that eq.(\ref{sfinbou}) is satisfied if we
impose
\beq
\label {finbou}
(\partial_\sigma X_\alpha +{\hat F}_{\beta\alpha}(X)
\partial_\tau X^\beta)|_{\sigma=0,\pi}=0
\eeq
along the world volume of the brane, while
the transverse coordinates still satisfy
\beq
\label {ffinbou}
X^i|_{\sigma=0}=y^i ~~~~~~~~~~i = p+1,..., d-1
\eeq
Translating these conditions in the closed channel we get the constraints
that the boundary state must satisfy to describe a {\dpb} with an external
field
\beq
\label {exfcl}
(\partial_\tau X_\alpha
+{\hat F}_{\beta\alpha} (X)\partial_\sigma X^\beta)|_{\tau=0}|B_X \rangle =0
\eeq
\beq
\label {fexfcl}
(X^i|_{\tau=0}-y^i)|B_X \rangle =0  ~~~~~~~~~~i = p+1,..., D-1
\eeq
%and the corresponding one for $\tau=T$ for the other brane.
Substituting in the previous equation the mode expansion 
we get the overlap conditions in terms of the oscillators.
Since they can be easily solved  only in the case of a constant 
$F_{\mu\nu}$ in the following we limit ourselves to a constant 
field~\footnote{A boundary state with non constant gauge field has been 
recently considered~\cite{HASHIMOTO}.}. 
In this case eqs.(\ref{exfcl}) and (\ref{fexfcl}) become:
\beq
\left\{
\begin {array}{l}
{\hat p}^\beta |B_X \rangle =0~~~;~~~{\hat q}^i |B_X \rangle =y^i|B_X \rangle \\
\left\{ (\one + {\hat{F}})^{\alpha}_{\,\,\,\beta} \alpha_{n}^{\beta} +
(\one - {\hat{F}})^{\alpha}_{\,\,\,\beta} {\widetilde{\alpha}}_{-n}^{\beta}
\right\}|B_X \rangle =0  \hspace{2cm} \,\,\, ,\\
(\alpha^i_{n}-\widetilde\alpha^i_{-n})|B_X \rangle =0
\end{array}\right.
\label {overexf}
\eeq
with $n \neq 0$.
They are satisfied by the following boundary  state
\begin{equation}
\label{bs5}
\ket{B_X} =N_p(F) \delta^{(d_\bot)}(\hat q - y)
\exp\biggl[-\sum_{n=1}^\infty \frac{1}{n}\,
\a_{-n}\cdot M \cdot
\widetilde \a_{-n}\biggr]\,
|0\rangle_\alpha|0\rangle_{\widetilde\alpha}|p=0\rangle~~,
\end{equation}
where
\begin{equation}
M^{\mu}\,\,\,_{\nu} \equiv ([(\one - {\hat{F}})(\one + 
{\hat{F}})^{-1}]^{\alpha}\,\,\,_{ \beta}
; - \delta^{i}\,\,\,_{j})
\label{modi4}
\eeq
and $N_p(F)$ is a normalization factor depending on the external field
that we will determine later.
Proceeding in an analogous way we can get the overlap conditions for the 
fermionic oscillators 
\beq
\left\{ (\one + {\hat{F}})^{\alpha}_{\,\,\,\beta} \psi_{t}^{\beta} - i \eta
(\one - {\hat{F}})^{\alpha}_{\,\,\,\beta} {\widetilde{\psi}}_{-t}^{\beta}
\right\}|B_{\psi} , \eta \rangle =0
\label{bosove2}
\eeq
for any half-integer [integer] value of $t$ for the NS [R] sector.
The boundary state satisfying the previous conditions is equal to:
\begin{equation}
\label{bs7}
\ket{B_\psi,\eta}_{\rm NS} = (-i) \exp\biggl[\ii\eta\sum_{t=1/2}^\infty
\psi_{-t}\cdot M \cdot \widetilde \psi_{-t}\biggr]
\,\ket{0}~~,
\end{equation}
for the NS sector and
\begin{equation}
\label{bs9}
\ket{B_\psi,\eta}_\R =-\exp\biggl[\ii\eta\sum_{t=1}^\infty
\psi_{-t}\cdot M \cdot \widetilde\psi_{-t}\biggr]
\,\ket{B_\psi,\eta}_\R^{(0)}~~,
\end{equation}
for the R sector, where the superscript $^{(0)}$ denotes the zero-mode
contribution that is equal to:
\beq
\label{bsr0}
\ket{B_\psi,\eta}_\R^{(0)} =
k(F)
{\cal M}_{AB}^{(\eta)}\,\ket{A} \ket{\widetilde B}~~,
\eeq
$k(F)$ is a normalization constant to be determined and
\begin{equation}
\label{bs14}
{\cal M}^{(\eta)} = C\Gamma^0\Gamma^{1}\ldots
\Gamma^{p} \,\left(
\frac{1+\ii\eta\Gamma_{11}}{1+\ii\eta}\right)U~~,
\end{equation}
$C$ is the charge conjugation matrix and  $U$ is equal to
\beq
U = ; {\rm e}^{-1/2 {\hat{F}}^{\alpha \beta} \Gamma_{\alpha}
\Gamma_{\beta}} ;
\label{umat}
\eeq
where the symbol $; \hspace{.5cm} ;$ means that we have to expand the
exponential and then antisymmetrize the indices of the $\Gamma$-matrices.
The boundary state for the matter fields is then obtained by
inserting in the first equation in (\ref{bounda4}) the
bosonic boundary state in eq.(\ref{bs5}) and the fermionic one given in
eq.(\ref{bs7}) for the NS sector and  in eq.(\ref{bs9}) for the R sector.
The complete boundary state is finally obtained by performing the
appropriate GSO-projections~\footnote{See eqs.(\ref{gso}) and  
Ref.~\cite{island} for more details.} 
on the state given in eq.(\ref{bounda3}) where the ghost contribution is
unchanged with respect to the case with $F=0$.

Analogously we get also the conjugate boundary state 
\begin{equation}
\label{bs1001}
\bra{B_X} = \langle p=0|\langle 0|_\alpha\langle 0|_{\widetilde\alpha}
\exp\biggl[-\sum_{n=1}^\infty \frac{1}{n}\, \a_{n}\cdot M\cdot
\widetilde \a_{n} \biggr]~~,
\end{equation}
for the bosonic coordinate,
\begin{equation}
\label{bs1011}
\bra{B_\psi,\eta}_{\rm NS} =  \ii\,\bra{0}\,
\exp\biggl[ \ii\,\eta\sum_{t=1/2}^\infty
\psi_{t}\cdot M\cdot\widetilde \psi_{t}\biggr]
\end{equation}
for the NS-NS sector, and
\begin{equation}
\label{bs1021}
\bra{B_\psi,\eta}_{\rm R} = - \bra{B,\eta}_{\rm R}^{(0)}
\exp\biggl[ \ii\,\eta\sum_{t=1}^\infty
\psi_{t} \cdot M\cdot\widetilde \psi_{t}\biggr]
\end{equation}
for the R-R sector, where
\beq
\bra{B,\eta}_{\rm R}^{(0)} =  (-1)^p k(\hat F) \bra{A} \,\bra{\widetilde B}
\left( C\Gamma^0\Gamma^{1}\ldots \Gamma^{p} \,
\frac{1+\ii\eta\Gamma_{11}}{1-\ii\eta}\, U\right)_{AB}~~
\label{brazemo}
\eeq
with $U$ given in eq.(\ref{umat}).

The coupling of a {\dpb} with the massless fields of the closed superstring
can be obtained by saturating the boundary state with the states 
corresponding to those fields. In the following we will show that the 
structure of those couplings is the same as that 
obtained from the DBI action and
actually the comparison with what comes from the DBI action allows us to
fix the normalization constants $N_p(F)$ and $k(F)$ appearing
in the boundary state. We want to stress, however, that the normalization
constants $N_p(F)$ and $k(F)$ could also be independently determined by 
requiring that the interaction between two branes be the same if we compute 
it in the open or in the closed string channel.

The coupling of a D$p$-brane with a specific massless field $\Psi$
can be computed by saturating the boundary state with the
corresponding field $\langle \Psi | $ ($\langle \Psi | $ can be
$\langle \Psi_h|,\langle \Psi_B|, \langle \Psi_\phi|$
corresponding respectively to the graviton, antisymmetric tensor and dilaton
or $\langle C_{(n)} |$ corresponding to a R-R state). 
By proceeding in this way we get the following couplings:
\begin{eqnarray}
J_{\phi} &\equiv & \frac{1}{2\sqrt{2}}\,J^{\mu\nu}\,
\left( \eta_{\mu\nu}-k_{\mu} \ell_{\nu} -
k_{\nu}\ell_{\mu}\right) \,\phi
\nonumber \\
&=&\frac{N_p(F)}{2 \sqrt{2}}\, V_{p+1}\,\,T_p ~
\left[ \frac{d-2p-4}{2}
+ {\rm Tr} \left( {\hat{F}} (\one + {\hat{F}})^{-1} \right) \right]
\,\phi~~;
\label{adil}
\end{eqnarray}
for the dilaton,
\beq
J_h \equiv J^{\mu \nu} \,h_{\mu\nu} =-  N_p(F) \,V_{p+1}\,T_p
~
\Big[(\eta + {\hat{F}})^{-1}\Big]^{\alpha \beta} h_{\beta\alpha}
\label{agra}
\eeq
for the graviton,
\begin{eqnarray}
J_{B} &\equiv & \frac{1}{\sqrt{2}}\, J^{\mu \nu} \,B_{\mu\nu} =
-\frac{N_p(F)}{2\sqrt{2}} \,V_{p+1} \,T_p
~\Big[(\eta-{\hat{F}})(\eta + {\hat{F}})^{-1} \Big]^{\alpha \beta}B_{\beta
\alpha}
\nonumber \\
&=&
-{\sqrt{2}}N_p(F)\,V_{p+1}\,\frac{T_p}{2}
~\Big[(\eta + {\hat{F}})^{-1} \Big]^{\alpha \beta}B_{\beta \alpha}
\label{aas}
\end{eqnarray}
for the NS-NS $2$-form potential and
\[
J_{C_{n}} \equiv \langle C_{(n)} \ket{B}_{R} =
-\frac{C_{\mu_1 \dots \mu_{n}}}{16 \sqrt{2}(n)! } V_{p+1}
N_p(F)k(F) \frac{T_p}{2} (1-(-1)^{p+n})
\]
\beq
Tr \left( \Gamma^{\mu_{n} \dots \mu_1} \Gamma^0
\dots \Gamma^p
 ;{\rm e}^{ - 1/2  {\hat{F}}_{\alpha \beta} \Gamma^{\alpha} \Gamma^{\beta}}
;\right)
\label{wnplus1}
\eeq
for the R-R states. The details of these calculations are given in 
Ref.~\cite{antone}.
The trace in eq.(\ref{wnplus1}) can be easily computed by expanding the
exponential term. The first term of the expansion of the exponential
gives the coupling of the boundary state with a $(p+1)$-form potential
of the R-R sector and is given by
\begin{equation}
J_{C_{(p+1)}} = \frac{\sqrt{2}\,T_p\,N_p(F)k(F)}{(p+1)!}\,
V_{p+1}\,C_{\alpha_0\ldots\alpha_p}\,\varepsilon^{\alpha_0\ldots\alpha_p}
\label{fite45}
\end{equation}
where we have used that for $d=10$ the $\Gamma$ matrices are $32\times 32$
dimensional matrices, and thus
$Tr( \one) = 32$ and that only the term with $n= p+1$
gives a non-vanishing contribution. Here
$\varepsilon^{\alpha_0\ldots\alpha_p}$ indicates the completely
antisymmetric tensor on the D brane world-volume~\footnote{Our convention
is that $\varepsilon^{0\ldots p}=-\varepsilon_{0\ldots p}=1$.}.
{F}rom \eq{fite45} we can immediately
deduce that the charge of a D$p$-brane with respect to
the R-R potential $C_{(p+1)}$ is
\beq
{\sqrt 2}T_p\,N_p(F)k(F)
\label{cou89}
\eeq
The next term in the expansion of the exponential
of \eq{wnplus1}
yields the coupling of the D$p$ brane with a
$(p-1)$-form potential which is given by
\begin{equation}
J_{C_{(p-1)}}=
\frac{{\sqrt 2}N_p(F)k(F)}{(p-1)!}\,
V_{p+1}\,\frac{T_p}{2}\,C_{\alpha_0\ldots\alpha_{p-2}}\,
{\hat F}_{\alpha_{p-1}\alpha_p}\,
\varepsilon^{\alpha_0\ldots\alpha_p}~~.
\label{sete56}
\end{equation}
where we have explicitly used 
the fact that only the term with $n=p-1$ gives a non-vanishing
contribution.
By proceeding in the same way, one can easily evaluate also the
higher order terms generated by the exponential
which describe the interactions of the D-brane
with potential forms of lower degree.
All these couplings can be encoded in the following term
\begin{equation}
\sum_{n=0}^{\ell_{\rm max}} \langle C_{(n)}|B\rangle _R=
{\sqrt 2}T_p\,N_p(F)k(F)\,\int_{V_{p+1}}\left[
\sum_{\ell=0}^{\ell_{\rm max}}
C_{(p+1-2\ell)}\,\wedge\,
{\rm e}^{\hat F}\right]_{p+1}
\label{WZte}
\end{equation}
where
$\ell_{\rm max}$ is
$p/2$ for the type IIA string and $\frac{p+1}{2}$ for the type
IIB string. We have defined
${\hat F} = \frac{1}{2} \,{\hat F}_{\a \b}\,
d\xi^{\a} \wedge d\xi^{\b}~$, while
\begin{equation}
C_{(n)} = \frac{1}{n!}\,C_{\alpha_1\ldots\alpha_n}\,d\xi^{\alpha_1}
\,\wedge\ldots\wedge\,d\xi^{\alpha_n}
\label{potn}
\end{equation}
$C_{\alpha_1\ldots\alpha_n}$ is the pullback of the $n$-form potential
 on the D-brane world-volume. The square bracket
in \eq{WZte} means that in expanding the exponential form
one has to pick up only the terms of total degree $(p+1)$,
which are then integrated over the $(p+1)$-dimensional world-volume.

The couplings of a Dp-brane with the massless states of closed superstring
that have been extracted from the boundary state can be compared with those
computed from its  low energy effective action.
The low energy effective action describing a {\dpb}
consists of a sum of two terms. The first one is the so called
Dirac-Born-Infeld
action, which in the string frame is given by
\beq
S_{DBI} = - \frac{T_{p}}{\kappa} \int_{V_{p+1}} d^{p+1} \xi
~{\rm e}^{-\varphi}
\sqrt{- \det\left[ {G}_{\alpha \beta} + {\cal B}_{\alpha \beta} +
\hat F_{\alpha \beta} \right] }~~.
\label{borninfe}
\eeq
where we are considering only its bosonic part, $2 \kappa^{2} =
(2\pi)^7 (\alpha ')^4 g_s^2$, ($g_s$ being the string
coupling) and $G_{\alpha \beta}$ and
${\cal B}_{\alpha \beta}$ are respectively the pullbacks of the space-time
metric and of the NS-NS antisymmetric tensor on the D-brane
world volume:
\beq
{G}_{\alpha \beta} = \partial_{\alpha} X^{\mu} \partial_{\beta} X^{\nu}
{G}_{\mu \nu} (X) \hspace{2cm} {\cal B}_{\alpha \beta} = \partial_{\alpha}
X^{\mu}
\partial_{\beta} X^{\nu} {\cal B}_{\mu \nu} (X)
\label{pulbac}
\eeq
Notice that the action in eq.(\ref{borninfe}) is
written using the string metric $G_{\mu \nu}$, and should be considered together
with the gravitational bulk action in the string frame
\[
S_{bulk} = \frac{1}{2\kappa^{2}} \int d^{10} x \sqrt{ - {G}} \left\{
~{\rm e}^{- 2 \varphi} \left[ R({G}) + 4 ( \nabla \varphi )^2
-\frac{1}{12} (H_3 )^2 \right] - \right.
\]
\beq
\left. - \sum_n \frac{1}{2\cdot (n+1)!} (F_{(n+1)} )^2  \right\} ~~.
\label{bulk}
\eeq
where $H_3$ [$F_{n+1}$] is the field strenght corresponding to the NS-NS [R-R]
$2$-form [$n$-form]  potential. 
In order to compare the couplings described by this action with the ones
obtained from the boundary state, it is first necessary to rewrite $S_{DBI}$
in the Einstein frame. In fact, like any string amplitude
computed with the operator formalism, also the couplings $J_h$, $J_\phi$
and $J_B$ are written in the Einstein frame. 
Furthermore, it is also convenient to introduce canonically normalized
fields
\beq
G_{\mu \nu} = {\rm e}^{\varphi/2} g_{\mu \nu}~~~,
~~~ \varphi =\sqrt{2} \kappa\, {\phi}~~~,~~~ {\cal {B}}_{\alpha \beta}  =
{\sqrt{2} \kappa}B_{\alpha \beta}{\rm e}^{\varphi/2}~~.
\label{einst}
\eeq
Inserting these fields in Eq. (\ref{borninfe}) we get
\beq
S_{DBI} = - \frac{T_{p}}{\kappa} \int d^{p+1} \xi~{\rm e}^{- \kappa {{\phi}}
(3- p )/(2\sqrt{2})}
\sqrt{- \det\left[ g_{\alpha \beta} + \sqrt{2} \kappa B_{\alpha \beta}
+ {\hat F}_{\alpha \beta }{\rm e}^{- \kappa {{\phi}}/ \sqrt{2}}
\right] }~~.
\label{borninfe2}
\eeq
By expanding the metric around the flat background as
\beq
g_{\mu \nu} = \eta_{\mu \nu} + {\tilde{h}}_{\mu \nu} = 
\eta_{\mu \nu} + 2 \kappa h_{\mu \nu}
\label{meexpa}
\eeq
and keeping only the terms
which are linear in the fields $h$, ${{\phi}}$
and $B$
one gets from eq.(\ref{borninfe2}) the following expression:
\begin{eqnarray}
S_{DBI}&\simeq& - ~T_p \int_{V_{p+1}} d^{p+1} \xi \sqrt{- \det \left[ \eta +
{\hat F}\right] }
~\Bigg\{
\Big[(\eta + {\hat{F}})^{-1}\Big]^{\alpha \beta} h_{\beta\alpha}
\label{expa}
\\
&-&\frac{1}{2\sqrt{2}}\,
\left[ 3 -p + {\rm Tr} \left( {\hat{F}} (\eta + {\hat{F}})^{-1} \right)
\right]
\,\phi
+ \frac{1}{\sqrt{2}} \Big[(\eta + {\hat{F}})^{-1}
\Big]^{\alpha \beta}B_{\beta \alpha}
\Bigg\}~~.
\nonumber
\end{eqnarray}
Comparing these couplings with those obtained in eqs.(\ref{adil}), 
(\ref{agra}) and (\ref{aas}) we can fix the normalization constant 
$N_p(F)$ to be
\beq
\label {normf}
N_p(F)=  \sqrt{- {\rm \det} (\eta + {\hat{F}})}
\eeq
Therefore the normalization factor of the boundary state
turns out to be proportional to  the Dirac-Born-Infeld Lagrangian.

The second term is the  Wess-Zumino action which is  given by
\beq
S_{WZ} =  \mu_p\int_{V_{p+1}} \sum_{n=0}^{\ell_{\rm max}} 
C_{(p+1-2n)} \wedge e^{{\hat{F}}}
\label{WZ}
\eeq
This has exactly the same structure as in eq.(\ref {WZte})
obtained by saturating  the boundary state with the R-R state.

Comparing eq.(\ref{WZ}) with eq.(\ref{WZte}) 
we can fix the other normalization constant to be
\beq
\label{kkappa}
k(F)=\frac{1}{\sqrt{{\rm det}(\eta+\hat F)}}
\eeq
and we can see that the coupling of a {\dpb} with the R-R (p+1)-form potential
is  given by
\beq
\label{corabr}
{\sqrt 2}T_p=\mu_p
\eeq
This is exactly the R-R charge carried by  the
$p$-brane classical solution of the low-energy string effective action 
in ten dimensions. 
In conclusion we have explicitly shown that by projecting
the boundary state $\ket{B}$ with an external field
onto the massless states of the closed string spectrum, one can
reconstruct the linear part of the
low-energy effective action of a D$p$ brane. This is the
sum of the Dirac-Born-Infeld part (\ref{expa})
and the Wess-Zumino term (\ref{WZte}) which are produced
respectively by the NS-NS and the R-R components of the boundary state.

For the sake of completeness we conclude this section by giving the 
asymptotic behaviour of the fields generated by a {\dpb}
with an arbitrary external field on it, determined by computing the quantity
\beq
\label{filflur}
\langle P_x|D|B\rangle
\eeq
where $D$ is the closed superstring propagator 
and $P_x$ are the projectors of the closed superstring massless sector
that can be found in Ref.~\cite{antone}.
For the massless NS-NS fields we get
\beq
\label{dilald}
\delta\phi=\frac{T_p V_{p+1}}{2{\sqrt 2}k^2_\perp}\sqrt{- \det
(\eta + {\hat{F}})}~
\left(\frac{(d-2p-4)}{2}+Tr\hat F(\one+\hat F)^{-1}\right),
\eeq
for the dilaton,
\[
\delta h^{\mu\nu}=-\frac{T_p V_{p+1}}{4k^2_\perp}\sqrt{- \det
(\eta + {\hat{F}})}~\left\{2\left(\eta+\hat F\right)^{-1~\alpha\beta}+
2\left(\eta+\hat F\right)^{-1~\beta\alpha}+
\right.\]
\beq
\label{grald}
\left.
+\frac{\eta^{\alpha\beta}}{2}
\left( -(1+p)+Tr\hat F(\one+\hat F)^{-1}\right),
\frac{\delta^{ij}}{2}
\left( -(1+p)+Tr\hat F(\one+\hat F)^{-1}\right)
\right\}
\eeq
for the graviton, and
\beq
\label{aald}
\delta B^{\alpha\beta}=-\frac{T_p V_{p+1}}{{\sqrt 2}k^2_\perp}\sqrt{- \det
(\eta + {\hat{F}})}~\left[
\left(\eta+\hat F\right)^{-1~\beta\alpha}-
~\left(\eta+\hat F\right)^{-1~\alpha\beta}\right]
\eeq
for the antisymmetric tensor.
Here, as usual $\alpha, \beta\in (0,...,p)$, $i,j\in(p+1,...,d-1)$ and
$k_\perp$ is the transverse momentum emitted from the brane.
For the massless R-R fields instead the long range fluctuaction around
the background values differs from the expressions
of the couplings contained in eq.(\ref{WZte}) simply for the inclusion of
the effect of the propagator, which generate a factor $1/k_\perp^2.$

From the previous expressions we can reconstruct the long distance
behaviour of the various fields in configuration space using the same
notation as in Refs.~\cite{island}.
For the metric tensor which is connected to the
graviton field by the relation in eq.(\ref{meexpa}) we get
\[
\delta {\tilde{h}}^{\mu\nu}(r)=-\frac{Q_p}{4r^{7-p}}\sqrt{- \det
(\eta + {\hat{F}})}~\left\{2\left(\eta+\hat F\right)^{-1~\alpha\beta}+
\left(\eta+\hat F\right)^{-1~\beta\alpha}+
\right.\]
\beq
\label{grar}
\left.
+\frac{\eta^{\alpha\beta}}{2}
\left( -(1+p)+Tr\hat F(\eta+\hat F)^{-1}\right),
\frac{\delta^{ij}}{2}
\left( -(1+p)+Tr\hat F(\eta+\hat F)^{-1}\right)
\right\}
\eeq
where $r$ is the radial coordinate in the transverse space.
For the dilaton field $\varphi$ which is connected to the canonically
normalized field by the second eq. in (\ref{einst})
we get
\beq
\label{dilar}
\delta\varphi(r)=\frac{Q_p}{4r^{7-p}}\sqrt{- \det
(\eta + {\hat{F}})}~
\left(3-p+Tr\hat F(\eta+\hat F)^{-1}\right),
\eeq
and for the antisymmetric tensor we get
\beq
\label{ald}
\delta {\cal{B}}^{\alpha\beta}(r)=-\frac{Q_p}{{2}r^{7-p}}\sqrt{- \det
(\eta + {\hat{F}})}~\left[
\left(\eta+\hat F\right)^{-1~\beta\alpha}-
~\left(\eta+\hat F\right)^{-1~\alpha\beta}\right]
\eeq
The field ${\cal{B}}$ is related to $B$ in eq.(\ref{aald}) by eq.(\ref{einst}).

\vskip 1.2cm
\section {T-duality and transverse excitations}
\label{traexci}
It is by now well known\footnote{See for instance sect. 4 of 
Ref.~\cite{island}} that, compactifying $d-p-1$ directions
and performing on them a T-duality transformation,  the abelian gauge
potential $A_\mu$ that lives on a D$9$-brane and that describes the massless 
excitations of an open string with NN boundary conditions in all directions
is transformed as follows: its longitudinal components
(those along the directions in which no
T-duality transformation is performed) give rise to a gauge field
$A_{\alpha}$ living on the world volume of a D$p$-brane, while its
transverse components
(those along the directions transformed by T-duality)
become the coordinates of the D$p$-brane. As the original gauge field, that
is related to the massless excitations of  an open string, corresponds to
the
collective coordinates of the D9-brane, so after the T-duality
transformation
both its longitudinal components and its transverse ones, that become the
coordinates of the Dp-brane, correspond to the collective coordinates of the
Dp-brane.

In the following we want to see how this conclusion can be
reached from the point of view of the boundary state.
To this purpose let us compare the overlap conditions for a boundary state
describing a {\dpb} with an external electric  field, that for simplicity we
take with
only one non vanishing component  $F^{0k}$ (here $k$  is one of the
longitudinal directions), with those for the boundary state describing a
D$(p-1)$-brane boosted in the transverse direction $k$ discussed in the
section \ref{boostrot}. In other words the direction $k$ is one of the 
longitudinal
directions of the D$p$-brane, while is one of the transverse directions
of the boosted D$(p-1)$-brane. Remember that a T-duality transformation in
a longitudinal direction transforms a {\dpb} into a D$(p-1)$-brane. 
Therefore if we
perform it on the {\dpb} that is described by the boundary state defined
through the overlap conditions given in the eq.(\ref{exfcl}) by remembering
that, as discussed in Ref.~\cite{island}, T-duality acts on the 
coordinate $X^k$ by
changing its $\tau$ derivative (Dirichlet-like boundary conditions in the
open channel) with minus its $\sigma$ derivative  (Neumann-like boundary 
conditions), we obtain that eq.(\ref{exfcl}) becomes:
\beq
\label {exfcl2}
(\partial_{\tau} X^0 - {\hat F}_k~^0\partial_\tau X^k)|_{\tau=0}|B\rangle
=0
\eeq
\beq
\label {exfcl3}
(\partial_\sigma X^k - {\hat F}(X)_{0}~^k\partial_\sigma X^0)|_{t=0}|B\rangle
=0
\eeq
By making the identification
\beq
\label{vf}
{\hat{F}}_0~^k={\hat{F}}_k~^0=- v
\eeq
the previous equations exactly reproduce eqs.(\ref{neuc3}) and
(\ref{dirc4bis}).
We can therefore conclude that the boundary state for a {\dpb} with an
external
electric field having only a non zero component in the direction $k$ is
equivalent, through a T-duality transformation along the direction $X^k,$ to
the boundary state of a  D$(p-1)$-brane boosted and delocalized along the 
transverse direction $k.$

Analogously it can  be shown that the boundary state for a {\dpb}
with an external magnetic field having $F_{\alpha k}$ as the only non vanishing
component  ($\alpha\neq 0$ and $k \neq 0$ are both longitudinal components)
is transformed
by a T-duality transformation on the coordinate $X^k$
to that of a D$(p-1)$-brane rotated in the
$(\alpha,k)$ plane and delocalized in the direction $X^k$. In this case in 
fact from eq.(\ref{exfcl}) we get
\beq
\label {frot1}
(\partial_\tau X^\alpha -{\hat F}_k~^\alpha\partial_\tau
X^k)|_{\tau=0}|B\rangle =0
\eeq
\beq
\label {frot2}
(\partial_\sigma X^k -{\hat F}_{\alpha}~^k\partial_\sigma
X^\alpha)|_{t=0}|B\rangle =0
\eeq
Then by making the identification
\beq
\label{frot3}
{\hat{F}}_\alpha~^k=-{\hat{F}}_k~^\alpha= \tan\omega
\eeq
the previous equations reproduce the overlap conditions for a D$(p-1)$-brane
which is rotated in the $(\alpha, k)$ plane and delocalized in the
direction $X^k$, that are given in eq.(\ref{rot22}) and in the equation
obtained by taking the $\sigma$ derivative of eq.(\ref{rot42}).

This result shows that after a T-duality transformation the T-dualized
components of the gauge field $U(1)$ are correctly reinterpreted as the
transverse coordinates of a D$(p-1)$-brane.
In fact if we consider a boost in the direction $k$ and start with $X^k=0$
we get
\beq
\label{bof}
\left\{
\begin{array}{l}
X'~^{0}=\frac{X^0}{\sqrt{1-v^2}}\\
X'~^{k}=\frac{vX^0}{\sqrt{1-v^2}}
\end{array}
\right.
\Rightarrow\partial_{0'}X'~^{k}=v
\eeq
Comparing eqs.(\ref{bof}) with eq.(\ref{vf})  we get
\beq
\label{pippo1}
{\hat{F}}_0~^k= - \partial_{0'}X'~^k  \Rightarrow \hat A^k \sim - (X')^k
\eeq
Analogously considering a rotation in the $(\alpha, k)$ plane
and starting with $X^k=0$ we get
\beq
\label{rof}
\left\{
\begin{array}{l}
X'~^\alpha={X^\alpha}\cos\omega\\
X'~^k=-X^\alpha\sin\omega
\end{array}
\right.
\Rightarrow \partial_{\alpha'}X'~^k=-\tan\omega
\eeq
that compared with eq. (\ref{frot3}) gives
\beq
\label{pippo2}
F_\alpha~^k=- \partial_{\alpha'}X'~^k \Rightarrow \hat A^k \sim - X^k
\eeq
showing that the transverse components of the gauge fields become the
transverse collective coordinates of the {\dpb}. Notice that in deriving both 
eqs.(\ref{pippo1}) and (\ref{pippo2}) we have assumed that 
$\partial_k A^0$ and $\partial_k A^{\alpha}$ are equal zero. This is a
consequence of the fact that the D$(p-1)$-brane is delocalized in the $k$
direction.

Since in the static gauge the $p+1$ coordinates of the world volume of a
D$p$-brane can be fixed to be $X^{\alpha} = \xi^{\alpha}$, a
D$p$-brane is described by the transverse coordinate $X^i$ and by the gauge
field $A_{\alpha}$ corresponding to a total of $10$ degrees of freedom
of which only $8$ are physical. Turning on a longitudinal gauge field in the
boundary state  as we have done
in eqs.(\ref{bs5}), (\ref{bs7}) and (\ref{bs9}) we have included
in the description of the Dp-brane only some of its excitations.
Following the previous discussion we can also include a dependence on its
transverse coordinates and therefore have a boundary state depending not
only
on a longitudinal gauge field, but also on the transverse coordinates of the
Dp-brane. From the previous discussion it is then immediately clear how this
can
be done: we must start from the boundary state of a D$9$-brane in which
$9-p$
directions are compactified and then perform a T-duality transformation
along
all of them. This can be easily done and we arrive at the following 
bosonic boundary state:
\beq
\label{fxbou}
|B_X\rangle =\frac{T_9}{2}\sqrt{-{\rm det}(\eta_{\alpha\beta}+
{\widetilde{F}}_{\alpha\beta})}~
e^{-\sum_{n=1}\frac{1}{n}\alpha_{-n}\cdot
M({\hat{F}},X^\perp)\cdot\widetilde\alpha_{-n}}
|0\rangle _\alpha |0\rangle _{\widetilde\alpha}|k=0\rangle
\eeq
where $M({\hat{F}},X^\perp)$ is the following $10\times 10$ matrix
\beq
\label{xfmat}
M({\hat{F}},X^\perp)_{\mu\nu} =
\left( \begin{array}{cc} [(1-\widetilde F)(1+\widetilde
F)^{-1}]^{\alpha\beta}~~
& ~~2\partial_\beta X^i(1+\widetilde F)^{-1~\alpha\beta} \\
2\partial_\beta X^i(1+\widetilde F)^{-1~\beta\alpha}~~
 &~~  -I+\partial_\alpha X^i\partial_\beta X^j
(1+\widetilde F)^{-1~\alpha\beta}
 \end{array} \right)
\eeq
and where we have defined
\beq
\label{tildef}
\widetilde F^{\alpha\beta}={\hat{F}}^{\alpha\beta}+
\partial^\alpha X^i\partial^\beta X_i
\eeq
The expression of the matrix $M({\hat{F}},X^\perp)$ has been determined by
observing that under a T-duality transformation
over $(9-p)$ coordinates
the exponential factor appearing in the boundary state changes as follows
\beq
\alpha^\mu_{-n} M_{\mu\nu}\widetilde\alpha^\nu_{-n}\rightarrow
\alpha^\alpha_{-n} M_{\alpha\beta}\widetilde\alpha^\beta_{-n}
-\alpha^\alpha_{-n} M_{\alpha j}\widetilde\alpha^j_{-n}+
\alpha^j_{-n} M_{j\alpha }\widetilde\alpha^\alpha_{-n}
-\alpha^i_{-n} M_{ij}\widetilde\alpha^j_{-n}
\label{ex564}
\eeq
where $\alpha,\beta\in (0,...p)$ and $i,j\in(p+1,....9-p)$. Then in terms of
the matrix appearing in the exponential factor of the boundary state
a T-duality transformation simply acts by changing the sign of all
the columns corresponding to the directions affected by the T-duality
transformation.

Saturating the previous boundary state with the various massless fields of
the bosonic sector of the closed superstring we reproduce the couplings
that one can read from the DBI action, for the case
$\partial^\alpha X^j\neq 0~\forall \alpha,j$ which are in fact given by
the following expression
\[
-\frac{T_p}{2}\int d^{p+1}\xi
\sqrt{-{\rm det}(\eta_{\alpha\beta}+\widetilde F_{\alpha\beta}})\left\{
\frac{\chi}{\sqrt 2}\left[ (p-3)-(1+\widetilde F)^{-1~\beta\delta}
\hat F_{\delta\beta}
\right] \right.  +
\]
\[
2(1+\widetilde F)^{-1~\beta\delta}\left(h_{\delta\beta}
+h_{ij}\partial_\delta X^i
\partial_\beta X^j+h_{\delta i}\partial_\beta X^i
+h_{\beta i}\partial_\delta X^i\right)
\]
\beq
\left. +\sqrt{2}
(1+\widetilde F)^{-1~\beta\delta}\left(B_{\delta\beta}
+B_{ij}\partial_\delta X^i
\partial_\beta X^j+B_
{\delta i}\partial_\beta X^i
+B_{\beta i}\partial_\delta X^i\right) \right\}
\label{bix}
\eeq

\vskip 1.5cm
\sect{(F,D$p$) bound states from the boundary state}
\label{fdp}
\vskip 0.5cm
In this section we are going to show that
the boundary state with an external {\it electric} field on it
can be used to obtain the long
distance behaviour of the various fields
describing the bound state (F, D$p$) formed by a
fundamental string and a D$p$-brane~\cite{antone}. This type of bound
state is a generalization of the dyonic string solution of Schwarz
\cite{SCHWARZ} and
has been recently discussed from the supergravity point of view
\cite{LU1,LU2,LU3}.
The (F, D$p$)
bound state can be obtained from a D$p$-brane by turning on
an {\it electric} field ${\hat F}$ on its world
volume.
With no loss in generality we can choose ${\hat F}$
to have non vanishing components only in the directions $X^0$ and $X^1$
so that it can be represented by the following $(p+1)\times (p+1)$
matrix
\begin{equation}
{\hat F}_{\alpha\beta} =
\left( \begin{array}{ccccc}
           0& -f & & &  \\
                       f & 0& & & \\
    & & 0 & &  \\
    & & &\ddots &  \\
    & & & & 0
    \end{array} \right)~~.
\label{effe3}
\end{equation}
Using this expression in \eq{modi4} one can easily see that the
longitudinal part of the matrix $M$ appearing in the boundary
state is given by
\begin{equation}
M_{\alpha\beta} =
\left( \begin{array}{ccccc}
           -\frac{1 + f^2}{1-f^2}  & \frac{2f}{1-f^2}& & &  \\
                       -\frac{2f}{1-f^2} & \frac{1+f^2}{1-f^2} & & & \\
    & & 1 & &  \\
    & & &\ddots &  \\
    & & & & 1
    \end{array} \right)
\label{emme3}
\end{equation}
while the transverse part of  $M$ is simply minus the identity
in the remaining $(9-p)$ directions.
Furthermore, using eq.(\ref{effe3}) one finds
\begin{equation}
- \det \left(\eta + {\hat{F}} \right) = 1 - f^2~~.
\label{emme31}
\end{equation}
Since we want to describe
configurations of branes with arbitrary R-R charge, we multiply the
entire boundary state by an overall factor of $x$. Later we
will see that the consistency of the entire construction will require
that $x$ be an integer, and also that the electric field strength $f$
cannot be arbitrary.

Let us now begin our analysis by studying the projection
in eq.(\ref{filflur}) in the NS-NS sector.
Using eqs. (\ref{dilald})-(\ref{aald}),
we find the long-distance
behavior of the NS-NS massless fields
generated by  the (F,D$p$) bound state.
For the dilaton we get
\begin{equation}
\label{dil4510}
\delta \phi = \mu_p \, \frac{V_{p+1} }{k_{\bot}^{2}}\,
x \frac{ f^2(p -5)+ (3-p) }{4\,\sqrt{1 - f^2}}
\end{equation}
For the antisymmetric
Kalb-Ramond field we find that,
since the matrix $M_{\mu\nu}$ is symmetric except in
the block corresponding to the $0$ and $1$ directions (see \eq{emme3}),
its only non-vanishing component
is  $B_{01}$ whose long-distance
behavior is given by
\begin{equation}
\delta B_{01}= {\mu_p}\,\frac{V_{p+1} }{k_{\bot}^{2}}\,
\frac{ x\,f }{\sqrt{1 - f^2}}~~.
\label{anti671}
\end{equation}
Finally, the components of the metric tensor
are
\begin{eqnarray}
\label{gra31}
\delta h_{00} &=& - \delta h_{11} =
{\mu_p}\,\frac{V_{p+1}}{k_{\bot}^2}\,x\,\frac{ f^2(p-1)+ (7-p)}
{8\sqrt{2}\sqrt{1-f^2}}~~,
\nonumber \\
\delta h_{22} &=& \ldots ~= \delta h_{pp} =
{\mu_p}\,\frac{V_{p+1}}{k_{\bot}^2}\,x\,\frac{ f^2(9-p)+ (p-7)}
{8\sqrt{2}\sqrt{1-f^2}}~~,
\\
\delta h_{p+1,p+1}&=& \ldots ~= \delta h_{99} =
{\mu_p}\,\frac{V_{p+1}}{k_{\bot}^2}\,x\,\frac{ f^2(1-p)+ (p+1)}
{8\sqrt{2}\sqrt{1-f^2}}~~.
\nonumber
\end{eqnarray}
Let us now turn to the R-R sector. In this case, after
the insertion of the closed string propagator, we have to
saturate the R-R boundary state  with the projectors on the
various R-R massless fields that can be found in Ref.~\cite{antone}.
Due to the structure of the R-R component
of the boundary state
describing the bound state (F,D$p$), it is not difficult
to realize that the only projectors that can give
a non vanishing result are those corresponding to a $(p+1)$-form and
to a $(p-1)$-form with all indices along the world-volume
directions. In particular, we find that the long distance
behavior of the $(p+1)$-form is given by
\begin{equation}
\delta C_{01\cdots p}
\equiv \bra{P^{(C)}_{01\cdots p}} \,D\,\ket{B}_{\rm R}
= - \mu_p \,\frac{V_{p+1}}{k_{\bot}^2}\,x~~.
\label{antiR2}
\end{equation}
Similarly, given our choice of the
external field, we find that the only non vanishing component
of the $(p-1)$-form emitted by the boundary state
is $C_{23\cdots p}$ whose long-distance behavior turns out to be
\begin{equation}
\delta C_{23\cdots p} \equiv \bra{P^{(C)}_{23\cdots p}}\,D \,\ket{B}_{\rm R}
=
- \mu_p\, \frac{ V_{p+1}}{k_{\bot}^{2}} \,x\,f~~.
\label{chi78}
\end{equation}
Notice that if $p=1$ this expression has to be interpreted as the
long-distance behavior of the R-R scalar which is usually denoted
by $\chi$.

In all our previous analysis, the two parameters $x$ and $f$
that appear in the boundary state seem to be arbitrary. However,
this is not so at a closer inspection. In fact,
they are strictly related
to the electric charges of the (F,D$p$) configuration under the Kalb-Ramond
field
and the R-R $(p+1)$-form potential.
It is well-known that these charges
must obey the Dirac quantization condition, {\it i.e.} they
must be integer multiples of the fundamental unit of (electric)
charge of a $p$-dimensional extended object $\mu_p$. In our notations
this quantization condition amounts to
impose that the coefficients of
$-\mu_p \,\frac{ V_{p+1}}{k_{\bot}^{2}}$ in Eqs. (\ref{anti671}) and
(\ref{antiR2}) be integer numbers. This implies that
\begin{equation}
x = n  ~~~~\mbox{and}~~~~ - \frac{x f}{\sqrt{1 - f^2}} = m
\label{xf}
\end{equation}
with $n$ and $m$ two integers. While the restriction on $x$
had to be expected from the very beginning because $x$ simply represents
the number of D$p$ branes (and hence of boundary states) that
form the bound state, the restriction on
the external field $f$ is less trivial.
In fact, from \eq{xf} we see that $f$
must be of the following form
\begin{equation}
f = - \frac{m}{\sqrt{n^2+m^2}}~~.
\label{xf3}
\end{equation}
This is precisely the same expression that appears in the
analysis of Ref.~\cite{SCHMID} on the dyonic string configurations, and is
also consistent with the results of Ref.~\cite{LU1,LU2,LU3}.

Using \eq{xf}, we can now rewrite the long distance
behavior of the massless fields produced by a (F,D$p$)
bound state in a more suggestive way. In doing so, we also
perform the Fourier transformation for $d=10$
 to work in configuration
space.
For later convenience, we also introduce the following
notations
\begin{equation}
\Delta_{m,n} = {m^2+n^2}
\label{Delta}
\end{equation}
Then, using \eq{dil4510} and assuming for the time being that the dilaton
has vanishing
vacuum expectation value, after some elementary steps, we obtain that the
long-distance behavior of the dilaton is
\begin{equation}
\varphi = \sqrt{2}\,\kappa\,\phi
\simeq  -\frac{n^2\,(p-3)+2\,m^2}{4\,\Delta_{m,n}}~\frac{Q_p}{r^{7-p}}~~,
\label{dilft}
\end{equation}
where $Q_p$ can be found in Ref.~\cite{antone}. 

Since we are going to compare our results with the
standard supergravity description of D-branes, we have reintroduced
the field $\varphi$ which differs from the canonically normalized
dilaton $\phi$ by a factor of $\sqrt{2}\,\kappa$ (see  eq.(\ref{einst})).
Similarly, recalling that $g_{\mu\nu}=\eta_{\mu\nu}+2\kappa\,h_{\mu\nu}$,
from \eq{gra31} we find
\begin{eqnarray}
g_{00} & = & -g_{11} \simeq -1 + \frac{6m^2-n^2\,(p-7)}{8\,\Delta_{m,n}}
~\frac{Q_p}{r^{7-p}}~~,
\nonumber \\
g_{22}&=&\ldots~=~g_{pp} \simeq
1 + \frac{2m^2-n^2\,(7-p)}{8\,\Delta_{m,n}}~\frac{Q_p}{r^{7-p}}~~,
\label{gft}
\\
g_{p+1,p+1} & = & \ldots~=~g_{99}
\simeq 1 + \frac{2m^2+n^2\,(p+1)}{8\,\Delta_{m,n}}
~\frac{Q_p}{r^{7-p}}~~.
\nonumber
\end{eqnarray}
Rescaling the Kalb-Ramond field by a factor of
$\sqrt{2}\,\kappa$ to obtain the standard
supergravity normalization and using eq.(\ref{anti671}), we easily
get
\begin{equation}
\label{aft}
{\cal B} = \sqrt{2}\kappa\,B \simeq
-\frac{m}{\Delta_{m,n}^{1/2}}\,\frac{Q_p}{r^{7-p}}
~dx^0\wedge dx^1~~.
\end{equation}
Finally, repeating the same steps for the R-R potentials
(\ref{antiR2}) and (\ref{chi78}) we find
\begin{equation}
\label{cpft}
{\cal C}_{(p+1)} = \sqrt{2}\kappa\,
C_{(p+1)}\simeq
-\frac{n}{\Delta_{m,n}^{1/2}}\,\frac{Q_p}{r^{7-p}}
~dx^0\wedge \ldots \wedge dx^p~~,
\end{equation}
and
\begin{equation}
\label{cp1ft}
{\cal C}_{(p-1)} = \sqrt{2}\kappa\,
C_{(p-1)}\simeq
\frac{m\,n}{\Delta_{m,n}}\,\frac{Q_p}{r^{7-p}}
~dx^2\wedge \ldots \wedge dx^p~~.
\end{equation}
Eqs. (\ref{dilft})-(\ref{cp1ft}) represent the leading
long-distance behavior of the massless fields emitted
by the (F,D$p$) bound state. Proceeding as in Ref.~\cite{antone} where
we have assumed  that the exact solution
can be written in terms of powers of the usual harmonic function
\begin{equation}
\label{hr}
H(r) = 1 + \frac{Q_p}{r^{7-p}}
\end{equation}
and also of the new harmonic function
\begin{equation}
\label{h1r}
H'(r) = 1 + \frac{n^2}{\Delta_{m,n}}~\frac{Q_p}{r^{7-p}}
\end{equation}
introduced in Ref.~\cite{LU2}, from
eqs.(\ref{dilft})-(\ref{cp1ft}) we can infer that
in the exact brane-solution corresponding to the (F, D$p$)
bound state the dilaton is
\begin{equation}
{\rm e}^{\varphi} = H^{-{1}/{2}}\,{H'}^{{(5-p)}/{4}}~~,
\label{dilfin}
\end{equation}
the metric is
\begin{eqnarray}
\label{metricfin}
ds^2 &=& H^{-{3}/{4}}\,{H'}^{{(p-1)}/{8}}
\left[-\left(dx^0\right)^2+\left(dx^1\right)^2
\right] \nonumber \\
&+&
H^{{1}/{4}}\,{H'}^{{(p-9)}/{8}}
\left[\left(dx^2\right)^2+\cdots+\left(dx^p\right)^2
\right] \nonumber \\
&+&
H^{{1}/{4}}\,{H'}^{{(p-1)}/{8}}
\left[\left(dx^{p+1}\right)^2+\cdots+\left(dx^9\right)^2
\right] ~~,
\end{eqnarray}
the Kalb-Ramond 2-form is
\begin{equation}
\label{krfin}
{\cal B} = \frac{m}{\Delta_{m,n}^{1/2}}~\left({H}^{-1}-1\right)
\,dx^0\wedge dx^1~~,
\end{equation}
and finally the R-R potentials are
\begin{equation}
\label{cpfin}
{\cal C}_{(p+1)} = \frac{n}{\Delta_{m,n}^{1/2}}~\left({H}^{-1}-1\right)
\,dx^0\wedge \cdots \wedge dx^p~~,
\end{equation}
and
\begin{equation}
\label{cp1fin}
{\cal C}_{(p-1)} = -\frac{m}{n}~\left({H'}^{-1}-1\right)
\,dx^2\wedge \cdots \wedge dx^p~~.
\end{equation}
In the case $p=1$, the last equation has to be replaced by
\begin{equation}
\label{axionfin}
\chi = -\frac{m}{n}~\left({H'}^{-1}-1\right)
\end{equation}
where $\chi$ is the R-R scalar field also called axion.

In writing this solution we have assumed that all fields except the
metric have vanishing asymptotic values. This explains why we have
subtracted the 1 in the last four equations. Our solution exactly
agrees with the one recently derived in Ref.~\cite{LU2} from
the supergravity point of view~\footnote{Actually, in comparing our
results with those of Ref.~\cite{LU2}, we find total agreement
except for the overall sign in the Kalb-Ramond 2-form. Our sign however
agrees with the dyonic string solution of Schwarz \cite{SCHWARZ}
when we put $p=1$.}. Moreover, eq.(\ref{axionfin}) can be shown to exactly
agree with the axion field of the dyonic string solution of Schwarz
\cite{SCHWARZ} in the case of vanishing asymptotic background values
for the scalars ($\varphi_0 = \chi_0 = 0$).
One can compute the vacuum amplitude between two boundary states at a 
distance $r$ from each other. This calculation has been  performed in 
Ref.~\cite{antone} where it has been found that the two branes do not 
interact.

We can therefore conclude that the boundary state with an external electric
field (eq.(\ref{effe3})), really provides
the complete conformal description of the BPS bound states formed
by fundamental strings and D$p$-branes.

\section{Compactified boundary state}
\label{compact}
\vskip 0.5cm
In this section we construct the boundary state describing a D$p$-brane 
which has all directions compactified on circles. From it, by decompactifying 
some directions, we can obtain the one in which only
certain directions are compactified. For the sake of simplicity we take 
radii all equal to $R$, but from the formulas that we will get, it will be
trivial to extend our results to arbitrary radii. Before we write the 
boundary state we first want to introduce a convenient notation. 
In the compactified
case it is convenient to introduce position and momentum operators
separately
for momentum and winding degrees of freedom requiring for them the following
commutation relations:
\beq
[ q_{w}^{\mu} , p_{w}^{\nu}] = i \eta^{\mu \nu}~~~~~~;~~~~~~
[ q_{n}^{\mu} , p_{n}^{\nu}] = i \eta^{\mu \nu}
\label{corel87}
\eeq
where the subscripts $n$ and $w$ correpond respectively to the momentum and
to the winding degrees of freedom and the other commutators are all
vanishing. By denoting with $| n^{\mu}, w^{\nu} \rangle $ an eigenstate of
the two "momentum" operators
\beq
p^{\rho}_{n}  | n, w  \rangle  = \frac{n^{\rho}}{R} | n,
w \rangle ~~~;~~~
p^{\rho}_{w}  | n, w \rangle  = \frac{w^{\rho}R}{\alpha'}
| n, w \rangle
\label{eige321}
\eeq
it is easy to convince oneself that the previous state can also be written
as
follows
\beq
| n, w \rangle  = e^{i q_n \cdot n/R} e^{i q_w \cdot w
R/\alpha'}
 | 0 , 0 \rangle ~~~;~~~
\label{nmuwnu}
\eeq
where $| 0, 0\rangle $ is the state with zero momentum and winding number.
The state in eq.(\ref{nmuwnu}) is normalized as:
\beq
\langle n, w| n' , w ' \rangle = \Phi\,\, \delta_{n n'} \delta_{w w'}
\label{norma31}
\eeq
where $\Phi$ is the "self-dual" volume that has the following properties:
\beq
\Phi = 2 \pi R~~~if~~~R \rightarrow \infty~~~~;~~~~
\Phi = \frac{2 \pi \alpha'}{R}~~~~if~~~R \rightarrow 0
\label{limi56}
\eeq
Let us use this formalism to write the boundary state for the compactified
case. In this case the part corresponding to the non-zero modes is
unchanged,
while the one corresponding to the zero modes of the bosonic coordinate
becomes~\cite{FRAU}:
\beq
|\Omega \rangle  = {\cal{N}}_p  \prod_{\alpha=0}^p \left[ \sum_{w^\alpha}
e^{i (q^{\alpha}_{w} - y^{\alpha}) w_{\alpha} R/\alpha'} \right]
\prod_{i= p+1}^{d-1} \left[ \sum_{n^i}
e^{i (q^{i}_{n} - y^{i}) n_{i}/R } \right] \ket{n=0 , w=0}
\label{comboun}
\eeq
where the parameters $y^{\alpha}$ and $y^i$ correspond respectively to Wilson
lines
turned on along the world volume of the brane and to the position of the
brane
in the transverse directions.

The previous boundary state satisfies the overlap conditions:
\beq
\left( {\em e}^{i (R/\alpha') q_{w}^{\alpha}}  - {\em e}^{ i (R/\alpha') 
y^{\alpha}} \right) |\Omega \rangle  = p_{n}^{\alpha}
|\Omega\rangle  =0~~~,~~~\alpha=0 \dots p 
\label{ove43}
\eeq
and
\beq
\left( {\em e}^{i q_{n}^{i}/R} - {\em e}^{ i y^{i}/R} \right) |\Omega 
\rangle  = p_{w}^{i} |\Omega\rangle  =0~~~,~~~i=p+1 \dots 9-p~~.
\label{ove45}
\eeq

The overall normalization can be determined by comparing the calculation of
the brane interaction done in the closed and open string channels. In this
way one gets the following relation~\cite{FRAU}:
\beq
{\cal{N}}_{p}^{2} \frac{\alpha'}{4} \Phi^d = \frac{VC_1}{2 \pi}
\label{norma71}
\eeq
where
\beq
V = (2 \pi R)^{p+1} \left(\frac{2 \pi \alpha'}{R} \right)^{d -p -1}~~~~~~;
~~~~~~~C_1 = (2 \pi)^{-d} ( 2 \alpha')^{-d/2}
\label{vc1}
\eeq
From eq.(\ref{norma71}) we get
\beq
{\cal{N}}_p = \sqrt{\frac{2V C_1}{\pi \alpha' (2 \pi R)^d}} (2 \pi R )^{d-p-1}
\left[ \left( \frac{2\pi R}{\Phi}\right)^{d/2}  (2\pi R)^{p+1-d} \right]
\label{np96}
\eeq
After some calculation it is easy to see that the part of
${\cal{N}}_p$ that is not contained in the square bracket just reproduces the
normalization of the boundary state in the uncompactified case and 
therefore we get:
\beq
{\cal{N}}_p = \frac{T_p}{2}
\left[ \left( \frac{2\pi R}{\Phi}\right)^{d/2}  (2\pi R)^{p+1-d} \right]
\label{np94}
\eeq
where $T_p$ is defined in eq.(\ref{bounda3}).
Let us  show  now that the previous normalization factor ${\cal{N}}_p$ 
reduces to
$T_p /2$ in the decompactified limit. In the decompactification limit  ($R
\rightarrow \infty$) one can easily check the following relations:
\beq
\sum_{w^\alpha} e^{i (q^{\alpha}_{w} - y^{\alpha}) w_{\alpha} R/\alpha'}
|0,0\rangle
\rightarrow |0,0\rangle
\label{inf519}
\eeq
and
\[
\sum_{n^i} e^{i (q^{i}_{n} - y^{i}) n_{i}/R/ }  \ket{0 , 0}
\rightarrow R \int dk e^{i(q_{n} - y )k } \ket{0,0} = 
\]
\beq
(2 \pi R)
\int \frac{dk}{2 \pi} e^{i(q_{n} - y )k } \ket{0,0} = 2 \pi R 
\delta ( q-y) \ket{0,0}
\label{inf520}
\eeq
where in the first relation we have taken into account that in the limit
$R \rightarrow \infty$ only the term with $w=0$ survives and in the second
relation we have substituted in the decompactification limit a sum with an
integral by introducing $k = n/R$. Using the two previous relations and the
first equation in (\ref{limi56}) it is easy to see that the normalization
factor reduces to $T_p /2$ that is the correct one in the decompactified
limit.
%Then, it is easy to check that the boundary state
%defined in this way
%correctly reproduces the vacuum amplitude for a
%wrapped D$p$-brane, normalization
%factors included.
If we decompactify only the time and the transverse directions the zero mode
contribution in eq.(\ref{comboun}) becomes
\beq
\frac{T_p}{2} \left(\frac{2\pi R}{\Phi} \right)^{p/2} \prod_{\alpha=1}^{p}
\left[\sum_{w^{\alpha}}{\rm e}^{i \theta^{\alpha} w^{\alpha}} 
\ket{n^{\alpha} =0 , w^{\alpha}} \right] \ket{k^0 =0}
\prod_{i=p+1}^{d-1} \left[ \delta (q^i - y^i ) \ket{ k^i =0}\right] 
\label{pcomp4}
\eeq
where $\theta = - y R/\alpha'$. 

\section{Stable non BPS states in type I theory}
\label{stable}

In this section, following the analysis of 
Sen~\cite{sen1,sen3,sen4,sen5,sen6}  and the approach of Frau et 
al.~\cite{LERDA}, we 
construct the boundary state 
corresponding to the stable non-BPS particle of type I theory that is related 
through the heterotic-type I duality to the state that belongs to the first 
excited level of the $SO(32)$ heterotic theory and that transforms according
to the spinor representation of $SO(32)$. Let us start by reminding 
some properties of the $SO(32)$ heterotic string and of its duality with
type I theory.
The heterotic string is a theory of closed oriented strings
with a local gauge invariance in ten dimensions.   It can be
considered as a combination of the bosonic string and the superstring. 
In fact it has a left sector, that is the same as the left sector of a
bosonic string in which $16$ of the $26$ coordinates are compactified
and  that is described by a lefthanded coordinate $X^{\rho} ( \tau  -
\sigma)$ where $\rho = 0 \dots 25$ and a righthanded one, that is the
same as the one of the superstring and that is described by the righthanded
coordinates ${\widetilde{X}}^{\mu} ( \tau + \sigma)$ and 
$ {\widetilde{\psi}}^{\mu} (\tau + \sigma )$ with $ \mu =0 \dots 9$.  
We can combine ${\widetilde{X}}^{\mu} ( \tau + \sigma)$ from the righthanded
sector with the first ten coordinates of the lefthanded sector $X^{\mu} 
( \tau - \sigma)$  to obtain the usual closed string coordinate:
\beq
X^{\mu} (\tau, \sigma ) = \frac{1}{2} \left({\widetilde{X}}^{\mu} 
( \tau + \sigma) + X^{\mu} ( \tau - \sigma) \right)~~~\mu= 0 \dots 9~~.
\label{rile}
\eeq
In addition we are left with $16$ compact righthanded coordinates $X^A ( \tau -
\sigma)$ with $A=1 \dots 16$ and of course with the lefthanded fermionic
coordinate $ {\widetilde{\psi}}^{\mu} (\tau + \sigma )$ with $ \mu =0 \dots 9$.

The mass spectrum of the heterotic string in the NS sector is determined by the
mass-shell conditions 
\begin{equation}
\label{hetL0}
\Big(L_0 -1\Big) \ket{\Psi}= \Big(\widetilde{L}_0 - \frac{1}{2}\Big)\ket{\Psi}
= 0~~,
\end{equation}
where we have used the values of the intercept of the bosonic theory 
($a=-1$) in the
left sector and of the NS superstring ($a=-1/2$) in the right sector. 
By expanding $\widetilde{L}_0$ in modes, one easily finds from the second 
equality in \eq{hetL0} that the mass $M$ of a state is given by
\begin{equation}
\label{hetmass}
M^2 = \frac{4}{\a'} \Big(\widetilde{N} -\shalf\Big)
\end{equation}
where $\widetilde{N}$ is the total number of right moving
oscillators.
From the first equality of \eq{hetL0} one can derive a generalized
level matching condition which relates 
$\widetilde{N}$ to the total number of left 
moving oscillators $N$ that are present in a given state.
This condition reads as follows
\begin{equation}
\label{lmc}
\widetilde{N} + \shalf  = N + \shalf \sum_A (p^A)^2~,
\end{equation}
where we have conveniently 
measured the internal momenta $p^A$ in units of $\sqrt{2\a'}$.
Additional restrictions come from the GSO projection that one has to perform 
in the right sector of the theory in order to have a consistent model. 
The GSO projection on NS states selects only 
half-integer occupation numbers $\widetilde N$. Since the
left occupation number $N$ is always integer as 
in the bosonic theory, in order to be able to satisfy \eq{lmc}, 
the internal momenta $p^A$ have to be quantized. 
In particular, the quantity $\sum_A(p^A)^2$ must be an even number. 
This condition implies that the internal coordinates $X^A$ 
must be compactified on an {\it even} 16-dimensional lattice.
The modular invariance of the one-loop partition function  requires
that this lattice be also {\it self-dual}. 
It can be shown that there exist only two 16-dimensional
lattices satisfying both these properties: the root lattice of $E_8\times E_8$, 
and a $Z_2$ sublattice of the weight lattice of $SO(32)$. 
Since we are interested in the heterotic theory with gauge group
$SO(32)$ here we will focus only on the second lattice which is denoted by 
$\Gamma_{16}$ and is defined by
\begin{equation}
\label{gamma16}
(n_1,\ldots,n_{16})\in \Gamma_{16}~~~\mbox{and}~~~
\Big(n_1+\shalf,\ldots,n_{16}+\shalf\Big)\in \Gamma_{16}
~~\Longleftrightarrow~~
\sum_i n_i \in 2 Z 
\end{equation}
The lowest states of the NS sector are massless and, because of 
eq.(\ref{hetmass}), 
all such states must have ${\widetilde N}=1/2$,
so that their right-moving part is simply 
${\widetilde\psi}_{-1/2}^i{\ket{\widetilde k}}$ where $i=2,...,9$ labels 
the directions transverse to the light-cone.
On the other hand, the level matching condition in eq.(\ref{lmc}) requires that 
\begin{equation}
N+\frac{1}{2}\sum_A\left(p^A\right)^2 = 1
\label{lmc1}
\end{equation}
This condition can be satisfied either by taking $N=1$ and $p^A=0$,
or by taking $N=0$ and the momenta $p^A$ to be of the 
form $P= (\pm 1,\pm 1,0,\ldots,0)$ (or any permutation thereof with
only both plus or both minus signs).
The first choice gives 16 states $\alpha_{-1}^A\ket{k;0}$,
while the second one contributes with $16 \times 15 \times 2 = 480$ states 
$\ket{k;P}$.
Altogether we have 496 massless states that carry a spacetime vector index
from the right-moving part and span the adjoint representation 
${\bf 496}$ of $SO(32)$. Those states correspond to the gauge fields of
$SO(32)$. At the massless level we have 64 more bosonic states 
which correspond to ${\widetilde N}=1/2$,
$N=1$ and $p^A=0$, and are given by
\begin{equation}
\label{state2}
\alpha_{-1}^i\ket{k;0}\,\otimes\,{\widetilde\psi}^j_{-1/2}
{\ket{\widetilde{k}}}
\end{equation}
where the indices $i,j$ run along the transverse directions $2,\cdots,9$.
These states are  singlets with respect to the gauge group but are 
space-time tensors. Decomposing them into irreducible components, we 
get a graviton, a dilaton and an antisymmetric two-index tensor. 
In conclusion, the bosonic massless states of the heterotic theory 
are in the following representations
\begin{equation}
({\bf 1};{\bf 1}) \oplus ({\bf 35};{\bf 1})
\oplus ({\bf 28};{\bf 1}) \oplus ({\bf 8};{\bf 496})
\label{repre1}
\end{equation}
where in each term the two labels refer to the Lorentz 
and gauge group respectively.
By analyzing the R sector, one finds an equal number of fermionic massless
states that complete the $N=1$ supersymmetric multiplets.

Let us now consider the first excited level of the NS sector 
that consists of states with $\widetilde{N}={3}/{2}$ and mass squared 
$M^2=4/\a'$. The states satisfying the condition $\widetilde{N}={3}/{2}$
are the following
\begin{eqnarray}
\label{1lr}
&&{\widetilde \psi}^i_{-3/2} 
{\ket{\widetilde{k}}} ~~\to~~\mbox{8 states}\\
\label{2lr}
&&{\widetilde\a}_{-1}^i{\widetilde\psi}^j_{-1/2} 
{\ket{\widetilde{k}}} ~~\to~~\mbox{64 states} \\
\label{3lr}
&& {\widetilde\psi}^i_{-1/2} {\widetilde\psi}^j_{-1/2} 
 {\widetilde\psi}^\ell_{-1/2} {\ket{\widetilde{k}}} ~~\to~~ \mbox{56 states}
\end{eqnarray}
By putting together the antisymmetric part of eq.(\ref{2lr}) together with 
the states
in eq.(\ref{3lr}) we obtain  a massive three-form transforming according to the
representation {\bf 84} of the Lorentz group, while the
remaining states transform together as a symmetric two-index tensor
in the representation {\bf 44}. 
The level matching condition (\ref{lmc}) imposes the constraint
\begin{equation}
N+\frac{1}{2}\sum_A\left(p^A\right)^2 = 2
\label{lmc2}
\end{equation}
There are 73,764 ways to satisfy this requirement! The complete list
of the corresponding states can be found for example on pag. 342 of 
Ref.~\cite{gsw} where it is shown that they transform as scalars, spinors,
second-rank antisymmetric tensors, fourth-rank antisymmetric tensors and
second-rank symmetric traceless tensors of $SO(32)$. 
Here we focus on the $2^{15}$ states 
that are obtained by taking in eq.(\ref{lmc2})
$N=0$ and momenta $p^A$ of the form
$(\pm\shalf,\pm\shalf,\ldots,\pm\shalf)$ with an even
number of plus signs. Notice that these momenta define a point
in the lattice $\Gamma_{16}$, since they satisfy
the second condition of eq.(\ref{gamma16}), and correspond
to the {\it spinor} representation of $SO(32)$. 
By combining these left modes with the 
right-moving ones of eqs. (\ref{1lr}) - (\ref{3lr}), 
we then obtain bosonic states transforming as
\begin{equation}
\label{trs}
({\bf 44};{\bf 2^{15}}) \oplus ({\bf 84};{\bf 2^{15}})~~.
\end{equation}
Analyzing the first excited level of the R sector, one can find
128 massive fermionic states which transform in the spinor representation
of $SO(32)$ and complete the $N=1$ supersymmetry multiplets.
Thus, altogether the spinors of $SO(32)$ appear with 256 different 
polarizations, 128 bosonic and 128 fermionic, corresponding to a {\it long}
multiplet of the $N=1$ supersymmetry algebra in ten dimensions. These states
are not BPS, but, nevertheless, stable. In fact, since there
are no spinors of $SO(32)$ at the massless level, they are the lighest states
with these quantum numbers and therefore cannot decay.

There is by now a strong evidence that the $SO(32)$ heterotic string is dual
to the type I theory~\cite{WIT}~\footnote{See also the second paper in 
Ref.~\cite{REVSEN}.}. They have the same spectrum of massless 
states, their
low-energy effective actions can be transformed into each other through the
following transformations on the metric and the dilaton:
\beq
G_{\mu \nu}^{H} = {\rm e}^{\phi_I} G_{\mu \nu}^{I}~~~~,~~~~
\phi_H = - \phi_I~~~.
\label{mediltra}
\eeq 
Since the second equation in (\ref{mediltra}) implies that the strong 
coupling limit of one theory is related to the weak coupling limit of the 
other theory, the fact that the perturbative massive spectra of the
two theories are totally different from each other is not in contradiction
with the fact that the two theories are non-perturbatively equivalent. 
For instance in the type I
theory all states transform according to the adjoint representation of the
gauge group $SO(32)$, while in the heterotic string the perturbative states
transform according to all the representations of $SO(32)$. But, if we have
in one of the two theory some state that cannot decay, then we expect it to 
be present in both theories at all values of the coupling constant. Identifying
such states in both theories is a check of duality. For instance an heterotic
string wrapped around a compact dimension breaks $1/2$ of supersymmetry and is
charged under the antisymmetric $2$-form of the gravitational multiplet, the
charge being simply its winding number. This is a BPS configuration and should
also appear in  type I theory. A natural candidate is the D string of type I
theory that is charged with respect to the R-R $2$-form of type I that 
corresponds
under the strong/weak duality to the antisymmetric $2$-form potential of the
gravitational multiplet of the heterotic string. But, since an heterotic string
is a BPS configuration with tension equal to
\beq
\tau_H = \frac{1}{2 \pi \alpha'}
\label{tauh}
\eeq 
it should match the tension of the D string of type I theory given by
\beq
\tau_{D1} = \frac{1}{2 \pi \alpha' g_I}
\label{taud1}
\eeq
In fact, if we use the metric relation in eq.(\ref{mediltra}), we see that
the two tensions in eqs.(\ref{tauh}) and (\ref{taud1}) transform into each
other. The identification between D string of type I and the fundamental
heterotic $SO(32)$ string can be tested at a deeper level by showing that the 
world sheet structure of a (wrapped) heterotic string is exactly reproduced
by the world-sheet dynamics on a (wrapped) D string~\cite{WIT}. Moreover, if 
the 
heterotic $SO(32)$ string theory is dual to the type I theory, then we must
be able to find in the latter one a description of the massive stable state
transforming according to the $2^{15}$ spinor representation of the heterotic
$SO(32)$ theory. In heterotic string units its mass is given by:
\beq
M_H = \frac{2}{\sqrt{\alpha '}} f (g_H )~~~,~~~~f(0) =1
\label{macca}
\eeq  
By going to type I units we expect its mass to be given by:
\beq
M_I = \frac{2}{\sqrt{\alpha '}} {\tilde{f}} (g_I )~~~~~,~~~~{\tilde{f}}(g)= 
f (1/g)
\label{mi}
\eeq
that in the weak coupling regime of type I theory becomes:
\beq
M_I = \frac{2}{\sqrt{\alpha '}} {\tilde{f}} (g_I \rightarrow 0 )
\label{mi0}
\eeq
where ${\tilde{f}} (0)$ cannot be determined  in type I perturbation theory.

From the heterotic $SO(32)$ point of view the stable state is just an excitation
of the fundamental string. Therefore in the type I theory we expect that a
D string should be involved in the description of this state. But a D string 
alone is not be sufficient because, on the one hand, it is a BPS 
configuration and on the other hand it is dual to the fundamental
heterotic string with winding number equal to $1$, while the stable state is 
not charged under the gravitational $2$-form of the heterotic string and 
therefore should be dual to a configuration of type I theory that is neutral 
under the
$2$-form RR potential of type I theory. Because of this the next simple 
possibility is that the stable state is a combination of a D string and an anti
D string:
\beq
|A > = | D1 > + | \overline{D1} >~~~~.
\label{d1antid1}
\eeq
But such a system is unstable because has tachyonic open string excitations. 
This can be easily seen by noticing that a change in sign in front of the
R-R spin structure, necessary in order to describe an anti D string in the
closed string channel, corresponds to a sign change of the $NS (-1)^{F}$ 
spin structure in
the open string channel. As a consequence the NS sector of the open strings
of the systems $1 -{\bar{1}}$ and  ${\bar{1}}-1$ contains only states that
are odd under $(-1)^F$ and their lowest states are  tachyons described in the
$-1$ picture by the states:
\beq
| k>_{-1} \lambda_{1 \bar{1}}~~~~~,~~~~~| k>_{-1} \lambda_{ \bar{1} 1}~~~~, 
\label{tachy}
\eeq
where we have introduced the following notation for the open string connecting
D strings and anti D strings:
\begin{eqnarray}
\mbox{for a 1-1 string}&\to& \lambda_{11} = \left(\begin{array}{cc}
 1 & 0 \\ 0 &  0
\end{array}\right)
\label{11} \\
\mbox{for a $\bar 1$-$\bar 1$ string}&\to& 
\lambda_{{\bar 1}{\bar 1}} = \left(\begin{array}{cc}
 0 & 0 \\ 0 &  1
\end{array}\right)
\label{1b1b} \\
\mbox{for a 1-$\bar 1$ string}&\to& \lambda_{1{\bar 1}} = 
\left(\begin{array}{cc}
 0 & 1 \\ 0 &  0
\end{array}\right)
\label{11b} \\
\mbox{for a $\bar 1$-1 string}&\to& \lambda_{{\bar 1}1} 
= \left(\begin{array}{cc}
 0 & 0 \\ 1 &  0
\end{array}\right)
\label{1b1}
\end{eqnarray}
The linear combination of the two tachyons in eq.(\ref{tachy}) that is even
under the $\Omega$ projection corresponding to the sum of the two states in
eq.(\ref{tachy}) will survive in the type I theory. In conclusion in type
I theory the state in eq.(\ref{d1antid1}) is unstable. But, even if we 
would find a way of eliminating the tachyon state, we will be left with the
problem that the state in eq.(\ref{d1antid1}) cannot represent the stable 
state of the heterotic string because both the D string and the anti D string
carry the quantum number of the spinor representation of $SO(32)$ and
therefore their bound state cannot transform itself under the spinor 
representation. By introducing, however, a Wilson line along the compactified 
direction
around which the anti D string is wrapped we can make it to transform as a
scalar of $SO(32)$. Therefore, if instead of the one in eq.(\ref{d1antid1})
we introduce the state
\beq
|A > = | D1 > + | \overline{D1} ' >~~~~,
\label{d1antid1'}
\eeq
where the prime denotes the Wilson line, it will have the correct quantum 
numbers for representing in type I theory the stable state of the heterotic 
$SO(32)$. This is the state proposed by Sen~\cite{sen4} to represent in type I
theory the stable non BPS perturbative massive state in eq.(\ref{trs}) in
the $SO(32)$ heterotic string.

Up to now we have not really used the fact that one direction has been 
compactified. In the following instead we want to consider the 
$D1/{\overline{D1}}$ system at the particular radius :
\beq
R = \sqrt{\frac{\alpha '}{2}}~~~~~.
\label{crira}
\eeq
This value is special for two reasons. From the mass-shell condition 
$L_0 -1/2 =0$ of the NS sector of the open type I theory we get the following
mass spectrum:
\beq
\alpha' M^2 = N - 1/2 + \frac{(n+1/2)^2}{R^2}~~~~~,
\label{specri}
\eeq
where $N$ is the oscillator number, $n$ is a integer and the quantity
$(n + 1/2)/R$ is the  Kaluza-Klein momentum obtained from the general
expression $p = n/R + \theta/(2 \pi R)$ for $\theta = \pi $ corresponding to
$Z_2$ Wilson lines. For the critical radius in eq.(\ref{crira}) the lighest 
string excitations, corresponding to the values $n=0, -1$, are massless and
not tachyonic. But since in the limit of
infinite radius they give rise to  tachyons, we will call them "tachyons"
also in the case we are considering. 

The second reason is that the conformal field theory generated by $X^1$ and
$\psi^1$, where the direction 1 corresponds to the compactified one, admits
several representations. One of them is the one obtained by just using $X^1$ 
and $\psi^1$. The other two can be obtained by fermionizing $X^1$
in terms of two additional fermions $\xi$ and $\eta$:
\beq
{\rm e}^{\pm i X^1 /\sqrt{2 \alpha'}} \simeq \frac{1}{\sqrt{2}} \left( \xi \pm
i \eta \right) ~~~~~,~~~~~\eta \xi \simeq i \partial X^1 /(2 \alpha ')~~~.
\label{xieta}
\eeq
The three fermions $\psi^1$, $\eta$ and $\xi$ are completely on equal
footing and can be regrouped in three  different ways. The first is the one
in which we combine $\xi$ and $\eta$ as in eq.(\ref{xieta}) and we represent
them in terms of the scalar field $X^1$ together with the fermionic field
$\psi^1$. The other two correspond in combining either $\xi$ and $\psi^1$ or
$\eta$ and $\psi^1$ in terms of respectively the scalar fields $\phi$ and 
$\phi'$ as follows
\beq
{\rm e}^{\pm i \phi^1 /\sqrt{2 \alpha'}} \simeq \frac{1}{\sqrt{2}} \left( \xi 
\pm i \psi^1 \right) ~~~~~,~~~~~\psi^1 \xi \simeq i \partial \phi /(2 \alpha ')
\label{psixi}
\eeq
and
\beq
{\rm e}^{\pm i \phi' /\sqrt{2 \alpha'}} \simeq \frac{1}{\sqrt{2}} \left( \eta 
\pm
i \psi^1 \right)~~~~~,~~~~~\psi^1 \eta \simeq i \partial \phi' /(2 \alpha ')~~~,
\label{psieta}
\eeq 
$\phi$ and $\phi'$ are bosonic fields compactified on a circle with radius
given in eq.(\ref{crira}). 

The advantage of using $\phi$ and $\eta$ instead of $X^1$ and $\psi^1$ is that 
in this case it is possible to explicitly encode the tachyonic background in 
the conformal field theory and to move the system from the unstable to a
stable situation. In order to show this let us start from the "tachyonic" states
in the $-1$ picture:
\begin{equation}
\label{tach+-} 
\ket{T_\pm} = ~\ex^{\pm{\ii\over \sqrt{2\a'}} X^1} 
\,\ket{0}_{-1}\otimes\sigma_1 \end{equation}
and look at the explicit form of their 
vertex operators ${\cal V}_{T_\pm}$ in the various representations.
Using the bosonization formulas (\ref{xieta}), (\ref{psixi}) and (\ref{psieta})
it is easy to see that 
\begin{equation}
\label{veq1}
{\cal V}_{T_\pm}^{(-1)} \,= \,\ex^{\pm{\ii\over \sqrt{2\a'}} X^1} 
\,\simeq \,
{1\over\sqrt{2}}( \xi \pm \ii \eta )   
\, \simeq \,
\left[\pm{\ii\over\sqrt{2}}\,\eta + 
\frac{1}{2}\left(\ex^{{\ii\over \sqrt{2\a'}}\phi}+\ex^{-{\ii\over 
\sqrt{2\a'}}\phi}\right)\right]
~~,
\end{equation}
where for simplicity we have understood the superghost part
and the Chan-Paton factor given by the Pauli matrix $\sigma_1$. From these 
relations
we immediately realize that the states $\ket{T_\pm}$ in eq.(\ref{tach+-})
can also be written either as 
\begin{equation}
\label{seq1}
\ket{T_{\pm}} =
{1\over\sqrt{2}}( \xi_{-{1\over 2}} \pm \ii \,\eta_{-{1\over 2}})
\ket{0}_{-1} \otimes\sigma_1 ~~,
\end{equation}
or as
\begin{equation}
\ket{T_{\pm}} = \left[\pm{\ii\over\sqrt{2}}\,\eta_{-{1\over 
2}}\ket{0}_\phi +{1\over 2}
\left(\ket{+{1\over 2}}_\phi+\ket{-{1\over 2}}_\phi\right)
\right] \ket{0}_{-1}  \otimes\sigma_1~~,
\end{equation}
where we have denoted by $\ket{\ell}_{\phi}$
the vacuum of $\phi$ with momentum $\ell$.
In particular, in the latter representation
the combination 
\begin{equation}
\label{calt}
\ket{\cal T} 
\equiv \frac{1}{\sqrt{2}\,\ii}\Big(\ket{T_+}-\ket{T_-}\Big)= 
\eta_{-{1\over 2}} \ket{0}_\phi \ket{0}_{-1} \otimes\sigma_1
\end{equation}
is formally identical to a
massless vector state at zero momentum in the $-1$
picture with $\psi$ replaced
by $\eta$. This implies that the deformation induced by 
$\ket{\cal T}$ corresponding to change the vacuum expectation value of the
"tachyonic state" can be described by the introduction of Wilson lines, 
that, however, should not be confused  with the $Z_2$ ones introduced above. 
In the $0$ picture ${\cal V}_{T_{\pm}}$ becomes
\begin{equation}
\label{-1/0}
{\cal V}_{T_\pm}^{(-1)}=\ex^{\pm{\ii\over \sqrt{2\a'}}X^1} \to 
{\cal V}_{T_\pm}^{(0)}=\pm \ii\, \psi^1 \ex^{\pm{\ii\over 
\sqrt{2\a'}}X^1}~;
\end{equation}
and then, by using the bosonization
formulas in eqs.(\ref{xieta}), (\ref{psixi}), and (\ref{psieta}) 
we can easily obtain the $(\phi$, $\eta)$ description of
${\cal V}_{T_\pm}^{(0)}$: 
\beq
{\cal V}_{T_{\pm}}^{(0)} = \pm i \psi^1 ( \xi  \pm i \eta )\sqrt{\alpha'}=
\mp \left( \partial \phi  \pm i \partial \phi' \right)
\label{vt+-}
\eeq
that implies
\begin{equation}
\label{vat0}
{\cal V}_{\cal T}^{(0)} = {\ii \over \sqrt{2\a'}}\,\partial \phi\,
\otimes\sigma^1~~.
\end{equation}
This is identical to the vertex of the usual gauge boson at zero 
momentum, where $\phi$ plays the role of the coordinate $X$. 
From this equation, it is clear that ${\cal V}_{\cal T}^{(0)}$ represents
a marginal operator which can be used to modify the 
theory. In particular we can modify the theory by introducing Wilson lines along
$\phi$ parametrized by
\begin{equation}
\label{wilsa}
W(\theta) = \frac{1}{2} Tr \left(\ex^{{\rm i} {\theta\over 2\sqrt{2\a'}}\oint 
d\sigma \, \partial_\sigma \phi \,\otimes\,\sigma^1} \right)
\end{equation}
where in a qualitative sense, the constant $\theta$ is equivalent to the 
tachyon vacuum expectation value since it multiplies the ``tachyon''
vertex operator ${\cal V}^{(0)}_{\cal T}$. By using the expansion:
\beq
\partial \phi = - 2 w \sqrt{\frac{\alpha'}{2}} + \dots~~~,
\label{phi78}
\eeq
where $\dots$ correspond to terms that do not give any contribution when we
perform the integral in eq.(\ref{wilsa}), we get
\beq
W (\theta ) = \cos \left( \theta \pi w/2 \right) 
\label{wilsa2}
\eeq
where $w$ is the total winding number of the closed string state as seen by the
operator $W (\theta)$. 

Let us now construct the boundary state of type IIB theory 
corresponding to the stable state following very closely the procedure 
described in Ref.~\cite{LERDA}. In the language of the boundary state the 
proposal of Ref.~\cite{sen4} corresponding to eq.(\ref{d1antid1'}), is given 
by  the superposition  of the boundary states describing respectively the D 
string and the anti D string. To describe this system we introduce the 
following 
boundary states
\begin{eqnarray}
\ket{B, +}_{\rm NS} &\equiv& \ket{D1, +}_{\rm NS} + 
\ket{D1', +}_{\rm NS}
\label{b1ns}\\
\ket{B, +}_{\rm R} &\equiv& \ket{D1, +}_{\rm R} - 
\ket{D1', +}_{\rm R}
\label{b1r}
\end{eqnarray}
where the $'$ indicates the presence of the ${\bf Z}_2$ 
Wilson line. 
Note that the 
minus sign in \eq{b1r} accounts for the fact that one of the
two members of the pair is an anti D-string.
Using the explicit expressions for the boundary state, 
we have~\cite{LERDA}
\begin{eqnarray}
\ket{B,+}_{\rm NS} &=&
\frac{T_1}{2}\,\sqrt{\frac{2\pi R}{\Phi}}
\,\exp\biggl[-\!\sum_{n=1}^\infty \frac{1}{n}\,
\a_{-n}\cdot \hat{S}^{(1)}\cdot \tilde \a_{-n}\biggr]
\exp\biggl[ +\,\ii \!\sum_{r=1/2}^\infty
\psi_{-r}\cdot \hat{S}^{(1)}\cdot\tilde \psi_{-r}\biggr]
\nonumber \\
&&\exp\biggl[-\!\sum_{n=1}^\infty \frac{1}{n}\,
\a_{-n}\tilde \a_{-n}\biggr]
\exp\biggl[ +\,\ii\! \sum_{r=1/2}^\infty
\psi_{-r}\tilde\psi_{-r}\biggr] \ket{\Omega}_{\rm NS}
\label{d1ns}
\end{eqnarray}
where we have denoted by ${\hat S}^{(1)}$ the 
D-string $S$-matrix 
for all non-compact directions and have separately 
indicated in the 
second line
the contribution of the bosonic and fermionic 
non-zero modes of 
the compact direction ({\it i.e.} the modes of $X$, $\psi$ and
$\widetilde\psi$). Due to the presence of the ${\bf Z}_2$ 
Wilson line,
the vacuum $\ket{\Omega}_{\rm NS}$ is given by
\begin{eqnarray}
\ket{\Omega}_{\rm NS} &=& \delta^{(8)}(q^i)\ket{k^0=0}
\left(\sum_{w}\ket{0,w}
+\sum_{w}(-1)^w\ket{0,w}\right)\prod_{i=2}^9\ket{k^i=0}
\nonumber \\
&=&2 \,\delta^{(8)}(q^i)\ket{k^0=0}\sum_{w}\ket{0,2w}
\prod_{i=2}^9\ket{k^i=0}
\label{omegans}
\end{eqnarray}
where for simplicity we have set to zero the 
coordinates $y^i$ of the 
D-strings. Analogously, in the R-R sector we have
\begin{eqnarray}
\ket{B,+}_{\rm R} &=&
\frac{T_1}{2}\,\sqrt{\frac{2\pi R}{\Phi}}
\,\exp\biggl[-\!\sum_{n=1}^\infty \frac{1}{n}\,
\a_{-n}\cdot \hat{S}^{(1)}\cdot \tilde \a_{-n}\biggr]
\exp\biggl[ +\,\ii \sum_{n=1}^\infty
\psi_{-n}\cdot \hat{S}^{(1)}\cdot\tilde \psi_{-n}\biggr]
\nonumber \\
&&\exp\biggl[-\!\sum_{n=1}^\infty \frac{1}{n}\,
\a_{-n}\tilde \a_{-n}\biggr]
\exp\biggl[ +\,\ii \sum_{n=1}^\infty
\psi_{-n}\tilde\psi_{-n}\biggr] \ket{D1,+}^{(0)}_{\rm R}\,
\ket{\Omega}_{\rm R}
\label{d1r}
\end{eqnarray}
where 
\beq
\ket{D1,+}^{(0)_R} = \left( C \Gamma^0 \Gamma^1 \frac{1 + i \Gamma_{11}}{1 +i}
\right)_{AB} |A> |{\tilde{B}}>
\label{vac42}
\eeq 
and
\begin{eqnarray}
\ket{\Omega}_{\rm R} &=& \delta^{(8)}(q^i)\ket{k^0=0}
\left(\sum_{w}\ket{0,w}
-\sum_{w}(-1)^w\ket{0,w}\right)\prod_{i=2}^9\ket{k^i=0}
\nonumber \\
&=&2 \,\delta^{(8)}(q^i)\ket{k^0=0}\sum_{w}\ket{0,2w+1}
\prod_{i=2}^9\ket{k^i=0}~~.
\label{omegar}
\end{eqnarray}
Let us now suppose that the radius $R$ is equal to the critical radius given 
in eq.(\ref{crira}) and rewrite the boundary state using a parametrization
along the compact direction in terms of the coordinate $\phi$ instead of $X^1$
as we have done in eqs.(\ref{d1ns}) and (\ref{d1r}). 
This will also allow the introduction of  $U(1)$
Wilson lines corresponding to a non vanishing vacuum expectation value for the
tachyon.  We are now in the position of writing the boundary state
which describes the D-string -- anti D-string pair in the
presence of a non vanishing tachyon v.e.v. This is given by
Eqs. (\ref{d1ns}) and (\ref{d1r}) with the oscillators
$\a_n$, $\tilde\a_n$, $\psi_r$, $\tilde\psi_r$ of the compact
direction replaced by $\phi_n$, $\tilde\phi_n$, $\eta_r$ and
$\tilde\eta_r$, and with a vacuum that carries an explicit
dependence on the parameter $\theta$ according to eq.(\ref{wilsa2}). 
In particular, in the NS-NS sector we have
\begin{eqnarray}
\ket{B(\theta),+}_{\rm NS} \!\! &=&\!\!
\frac{T_1}{2}\sqrt{\frac{2\pi R_c}{\Phi}}
\exp\biggl[-\!\sum_{n=1}^\infty \frac{1}{n}\,
\a_{-n}\cdot \hat{S}^{(1)}\cdot \tilde \a_{-n}\biggr]
\exp\biggl[ +\,\ii \!\sum_{r=1/2}^\infty
\psi_{-r}\cdot \hat{S}^{(1)}\cdot\tilde \psi_{-r}\biggr]
\nonumber \\
&&\!\exp\biggl[-\!\sum_{n=1}^\infty \frac{1}{n}\,
\phi_{-n}\tilde \phi_{-n}\biggr]
\exp\biggl[+\ii\! \sum_{r=1/2}^\infty
\eta_{-r}\tilde\eta_{-r}\biggr] 
\ket{\Omega(\theta)}_{\rm NS}
\label{hatbns}
\end{eqnarray}
where
\begin{equation}
\ket{ \Omega(\theta)}_{\rm NS} = 
2 \,\delta^{(8)}(q^i)\ket{k^0=0}
\sum_{w_\phi}\cos(\pi\theta 
w_\phi)\,\ket{0,2w_\phi}\prod_{i=2}^9\ket{k^i=0}~~.
\label{omegahatns}
\end{equation}
Analogously, in the R-R sector we have
\begin{eqnarray}
\ket{ B (\theta),+}_{\rm R} \!&=&\!
\frac{T_1}{2}\sqrt{\frac{2\pi R_c}{\Phi}}
\,\exp\biggl[-\!\sum_{n=1}^\infty \frac{1}{n}\,
\a_{-n}\cdot \hat{S}^{(1)}\cdot \tilde \a_{-n}\biggr]
\exp\biggl[ +\,\ii \!\sum_{n=1}^\infty
\psi_{-n}\cdot \hat{S}^{(1)}\cdot\tilde \psi_{-n}\biggr]
\nonumber \\
&&\exp\biggl[-\!\sum_{n=1}^\infty \frac{1}{n}\,
\phi_{-n}\tilde \phi_{-n}\biggr]
\exp\biggl[+\ii\! \sum_{n=1}^\infty
\eta_{-n}\tilde\eta_{-n}\biggr] \ket{D1,+}^{(0)}_{\rm R}\,
\ket{\Omega(\theta)}_{\rm R}
\label{hatbr}
\end{eqnarray}
where
\begin{equation}
\ket{\Omega(\theta)}_{\rm R} = 
2 \,\delta^{(8)}(q^i)\ket{k^0=0}\sum_{w_\phi}
\cos\left(\pi\theta(w_\phi+\frac{1}{2})\right)\,
\ket{0,2w_\phi+1}\prod_
{ i = 2 }^9\ket{k^i=0}~~. \label{omegahatr}
\end{equation}
Notice that at $\theta=0$ the boundary 
states (\ref{hatbns}) and 
(\ref{hatbr}) reduce to 
the original ones written in 
eqs.(\ref{d1ns}) and (\ref{d1r}). 
However the idea is that, for arbitrary values of $\theta$, only 
the boundary states
(\ref{hatbns}) and (\ref{hatbr}) must be used. 

In the computation of the interaction between  two boundary states of the 
type as in
eqs.(\ref{hatbns}) and (\ref{hatbr}) we must perform the GSO projection that 
acts on
the boundary state in terms of the variables $\phi$ and $\eta$ differently
than on the original variables $X^1$ and $\psi^1$. From eq.(\ref{psixi})
one can immediately read the action of $(-1)^{\tilde{F}}$ getting in the 
NS-NS sector 
\begin{equation}
%(-1)^F ~~:~~ \a_{n}^\mu \to \a_{n}^\mu~~,~~ 
%\psi_r^\mu \to -\psi_r^\mu
%~~,~~\phi_n \to -\phi_n~~,~~\eta_r\to\eta_r
%\nonumber \\
(-1)^{\widetilde F} ~~:~~ \widetilde\a_{n}^\mu \to 
\widetilde\a_{n}^\mu~~,~~ \widetilde\psi_r^\mu 
\to -\widetilde\psi_r^\mu
~~,~~\widetilde\phi_n \to 
-\widetilde\phi_n~~,~~\widetilde\eta_r\to
\widetilde\eta_r 
\label{-1f}
\end{equation}
and similarly for $(-1)^{F}$ on the left moving 
oscillators. Notice that the action of $(-1)^{{\tilde{F}}}$ on $\phi$
looks very much like T-duality because it amounts to change the 
relative sign between its left and right moving oscillators. 
It acts also on the zero modes as follows:
\beq
\sqrt{\frac{\alpha'}{2}}\left( \frac{n_{\phi}}{R} -  
\frac{w_{\phi} R}{\alpha'}\right) = n_{\phi} - \frac{w_{\phi}}{2} 
\longrightarrow - \left(n_{\phi} - \frac{w_{\phi}}{2} \right)
\label{ftra}
\eeq
implying that
\beq
n_{\phi} \rightarrow \frac{w_{\phi}}{2}~~~~,~~~~~ w_{\phi} 
\rightarrow 2 n_{\phi}
\label{ftra2}
\eeq
This implies
\begin{equation}
(-1)^{\widetilde F}~~:~~ \ket{n_\phi,w_\phi} ~\rightarrow
\ket{\frac{w_\phi}{2}, 2n_\phi}~~.
\label{nwexchange}
\end{equation}
Of course, similar considerations 
apply also for the boundary states in the R-R sector.
 Using the previous rules rules, one can easily see, for example, 
that
\begin{eqnarray}
%(-1)^F\,\ket{B(\theta),+}_{\rm NS}
%&=&
(-1)^{\widetilde F}\,\ket{B(\theta),+}_{\rm NS}
&=&-
T_1\,\sqrt{\frac{2\pi R_c}{\Phi}}
\,\exp\biggl[-\!\sum_{n=1}^\infty \frac{1}{n}\,
\a_{-n}\cdot \hat{S}^{(1)}\cdot \tilde \a_{-n}\biggr]
\nonumber \\
&&\exp\biggl[ -\,\ii \!\sum_{r=1/2}^\infty
\psi_{-r}\cdot \hat{S}^{(1)}\cdot\tilde \psi_{-r}\biggr]
\label{-1fb} \\
&&\exp\biggl[+\sum_{n=1}^\infty \frac{1}{n}\,
\phi_{-n}\tilde \phi_{-n}\biggr]
\exp\biggl[+\ii\!\sum_{r=1/2}^\infty
\eta_{-r}\tilde\eta_{-r}\biggr] \nonumber \\
&&\delta^{(8)}(q^i)\ket{k^0=0}
\sum_{w_\phi}\cos(\pi\theta 
w_\phi)\,\ket{w_\phi,0}\prod_{i=2}^9\ket{k^i=0}
~~.\nonumber 
\end{eqnarray}
Let us now compute the vacuum amplitude of the theory 
defined
on the world-volume of our D-string -- anti D-string pair. 
In the boundary state
formalism this amplitude is simply given by
\begin{equation}
{\cal A}(\theta) = \bra{B(\theta),+}\,P_{\rm GSO}\,
D\,\ket{B(\theta),+}
\label{ampl}
\end{equation}
where the GSO projection operator is
given in eqs.(\ref{gso}) and $D$ is the closed string propagator
\beq
D = \frac{\alpha'}{4\pi} \int \frac{d^2 z}{|z|^2} z^{L_0 -a } 
{\bar{z}}^{{\tilde{L}}_0 - a}
\label{d}
\eeq
with intercept $a_{\rm NS}=1/2$ in the NS-NS sector, and 
$a_{\rm R}=0$
in the R-R sector. Using the explicit expressions of the 
boundary states written above, and performing 
standard manipulations, one finds~\cite{LERDA}
\begin{equation}
{\cal A}_{\rm NS-NS}(\theta)
= \frac{VR_c}{2\pi \a '}
\int_{0}^\infty \!\frac{dt}{t^4}  \!\left[\!
\left(\sum_{w_\phi}\cos^2(\pi\theta w_\phi) ~q^{w_\phi^2}\right)
\!\frac{f_3^8(q)}{f_1^8(q)}
-\sqrt{2}\,\frac{f_4^7(q)\,f_3(q)}{f_1^7(q)
\,f_2(q)}
\right]
\label{vacampl}
%\end{eqnarray}
\end{equation}
and
\begin{equation}
{\cal A}_{\rm R-R}(\theta) = - \frac{VR_c}{2\pi \a '}
\int_{0}^\infty \frac{dt}{t^4} 
\left[\sum_{w_\phi}\cos^2\left(\pi\theta (w_\phi +\frac{1}{2})\right)
~q^{\left(w_\phi+\frac{1}{2}\right)^2}\right]
\frac{f_2^8(q)}{f_1^8(q)}
\label{vacampl1}
\end{equation}
where $V$ is the (infinite) length of the time direction 
%and 
%\begin{eqnarray}
%f_1(q) &=& q^{1 \over 12} \prod_{n=1}^{\infty} (1-q^{2n})
%~~~,~~~f_2(q) = \sqrt{2} q^{1 \over 12}
% \prod_{n=1}^{\infty} (1+q^{2n})\nonumber\\
%f_3(q) &=& q^{-{1 \over 24}} \prod_{n=1}^{\infty} (1+q^{2n-1})
%~~~,~~~
%f_4(q)= q^{-{1 \over 24}} \prod_{n=1}^{\infty} (1-q^{2n-1})
%\end{eqnarray}
%with $q={\rm e}^{-t}$.

It is interesting to observe that the contribution of the NS-NS$(-1)^F$ spin
structure  ({\it i.e.} the second term in \eq{vacampl}) 
does not depend 
on the tachyon v.e.v. $\theta$. This is a direct 
consequence of the
fact that this  spin structure arises from the 
overlap between
$\ket{B(\theta),+}_{\rm NS}$, 
whose vacuum contains states with only
winding numbers, and $(-1)^{\widetilde F}
\ket{B(\theta),+}_{\rm NS}$, whose vacuum instead 
contains states with only Kaluza-Klein numbers (see
eqs.(\ref{omegahatns}) and (\ref{-1fb})). Therefore, in the
NS-NS$(-1)^F$ spin structure there is no contribution from the
bosonic zero modes of the compact direction $\phi$, and hence
no dependence on the tachyon v.e.v. $\theta$. 

If one performs the modular transformation $t \to 1/t$, 
the entire amplitude ${\cal A}(\theta)$
can be interpreted as the one-loop vacuum energy of the open strings
living in the world-volume of the D-string -- anti D-string pair.

%After this modular transformation, one can explicitly 
%check that
%${\cal A}(\theta)$ in \eq{ampl} coincides with the  
%annulus amplitude that follows from the rules 
%described in Ref.~\cite{sen4} from the open
%string point of view (see Appendix A for some details). 
In particular, one sees that the $\theta$-independent
NS-NS$(-1)^F$ spin structure of the closed string 
channel goes into the
R spin structure of the open string channel, that 
indeed has been shown
in Ref.~\cite{sen4} to be independent of the tachyon v.e.v.
 
At $\theta=1$ a remarkable simplification 
occurs: the R-R contribution to ${\cal A}$
vanishes and the entire vacuum amplitude 
becomes 
\begin{eqnarray}
{\cal A}(\theta=1) &=& \frac{VR_c}{2\pi \a '}
\int_{0}^\infty \!\frac{dt}{t^4}  \!
\left[\left(\sum_{w_\phi} q^{w_\phi^2}\right)
\frac{f_3^8(q)}{f_1^8(q)}
-\sqrt{2}\,\frac{f_4^7(q)\,f_3(q)}{f_1^7(q)\,f_2(q)}
\right]
\nonumber \\
&=& \frac{V}{4\pi R_c}
\int_{0}^\infty \! \frac{dt}{t^4} 
\left(\sum_{w_\phi} q^{w_\phi^2}\right)
\left[\frac{f_3^8(q)}{f_1^8(q)}
-\frac{f_4^8(q)}{f_1^8(q)}\right]
\label{ampl1}
\end{eqnarray}
that is obtained using the following identities
\begin{equation}
f_2(q)\,f_3(q)\,f_4(q) = \sqrt{2}~~~~,~~~~f_1(q)\,
f_3^2(q)
=\sum_{n=-\infty}^{+\infty} q^{n^2}
\label{ident}
\end{equation}
Notice that with these 
manipulations we have
managed to reconstruct the typical combination of 
$f$-functions that is produced by the {\it usual} GSO projection 
of the NS-NS sector.
Thus, one is lead to think that a simpler 
underlying structure may actually
exist at $\theta=1$. In fact  the amplitude ${\cal A}(\theta=1)$
can be factorized in terms of a new boundary 
state according to~\cite{LERDA}
\begin{equation}
{\cal A}(\theta=1) = \bra{\widetilde B,+}\,P_{GSO}\,
D\,\ket{\widetilde B,+}
\label{ampl2}
\end{equation}
where
\begin{eqnarray}
\ket{\widetilde B,\pm} &=&
\frac{T_1}{2}\,\sqrt{\frac{\pi\a'}{R_c\Phi}}
\,\exp\biggl[-\!\sum_{n=1}^\infty \frac{1}{n}\,
\a_{-n}\cdot \hat{S}^{(1)}\cdot \tilde \a_{-n}\biggr]
\exp\biggl[ \pm\,\ii \!\sum_{r=1/2}^\infty
\psi_{-r}\cdot \hat{S}^{(1)}\cdot\tilde \psi_{-r}\biggr]
\nonumber \\
&&\!\exp\biggl[+\sum_{n=1}^\infty \frac{1}{n}\,
\a_{-n}\tilde \a_{-n}\biggr]
\exp\biggl[ \mp\ii\! \sum_{r=1/2}^\infty
\psi_{-r}\tilde\psi_{-r}\biggr] \ket{\widetilde\Omega}
\label{newbound}
\end{eqnarray}
with
\begin{equation}
\ket{\widetilde \Omega}= 
2 \,\delta^{(8)}(q^i)\ket{k^0=0}
\sum_{w}\ket{w,0}\prod_{i=2}^9\ket{k^i=0}~~.
\label{newomega}
\end{equation}
Of course, the simple factorization of a vacuum amplitude
does not allow to conclude that the two boundary states that have been used
are completely equivalent. However, a detailed analysis
of correlation functions shows that the new boundary state
$\ket{\widetilde B,+}$, which is written in 
terms of the 
original degrees of freedom for the compact 
direction ({\it i.e.}
$X$ and $\psi$), is equivalent to the boundary state
of \eq{hatbns} for $\theta=1$. For details see Ref.~\cite{LERDA}.

Based on these results, we can conclude that in order 
to describe the D-string -- anti D-string pair at $R=R_c$
in terms of $X$ and $\psi$, we have to use the original 
boundary states of eqs.(\ref{d1ns}) and (\ref{d1r})
if $\theta=0$, whereas we have to use 
the new boundary state of \eq{newbound}
if $\theta=1$.  Of course, at this particular 
value of $\theta$ there is no R-R sector, as we have 
explicitly shown. It is interesting to observe that
$\ket{B,+}_{\rm NS}$ and $\ket{\widetilde B,+}$ can
be related to each other by means of a
``generalized T-duality''. Indeed,
as is clear from eqs.(\ref{d1ns}) and (\ref{newbound}),
we may go from $\ket{B,+}_{\rm NS}$ to 
$\ket{\widetilde B,+}$ by changing the sign to the
right moving oscillators of the compact direction, and
consequently by changing the vacuum from $\ket{0,2w}$ to 
$\ket{w,0}$ since the radius of $X$ satisfies eq.(\ref{crira})
(see also eq.(\ref{nwexchange})).
Like the true T-duality, also this ``generalized
T-duality'' transforms a longitudinal direction into a 
transverse one, so that the new boundary state
$\ket{\widetilde B,+}$ describes a
D0-brane with a compact transverse direction.
However, unlike the true T-duality,
the ``generalized T-duality'' switches off the R-R sector.
This fact suggests that, more than a symmetry of the theory,
this ``generalized T-duality'' has to be regarded simply as
an effective way of implementing the change of
the tachyon v.e.v. from $\theta=0$ to $\theta=1$ on the 
original boundary states, which can be justified
by introducing the new fields $\phi$ and $\eta$
through the bosonization procedure as it has been done above~\cite{LERDA}.

Up to now we have worked at the critical radius given in eq.(\ref{crira}).
As pointed out in Ref.~\cite{LERDA} the decompactification limit is ill 
defined on  the original boundary states
($ \ket{B, +}_{NS}$ and  $\ket{B, +}_{R} $) because their vacuum contains 
only odd
or even winding numbers, but it is perfectly well defined on the new one
($\ket{{\tilde{B}}, +}_{NS}$) at $\theta =1$. In fact when $R \rightarrow 
\infty$ we get
\begin{eqnarray}
\ket{\widetilde\Omega} &=& 2\,
\delta^{(8)}(q^i)\,\ket{k^0=0}~2\pi R\int
\frac{dk^1}{2\pi}\ket{k^1}\,\prod_{i=2}^9\ket{k^i=0}
\nonumber \\
&=&4\pi R\,\delta^{(9)}(q^i)\,
\prod_{\mu=0}^9\ket{k^\mu=0}
\label{newomega0}
\end{eqnarray}
which corresponds to the vacuum structure of a $0$ brane.
Furthermore,
combining the factor
of $4\pi R$ from \eq{newomega0} with the prefactors of
$\ket{\widetilde B,+}$ in eq.(\ref{newbound}), we see that
the complete normalization of the boundary state 
becomes
\begin{equation}
\frac{T_1}{2}\,\sqrt{\frac{\pi\a'}{R\Phi}}~
4\pi R ~~~\longrightarrow ~~~
\frac{\sqrt{2}\,T_1\,(2\pi\sqrt{\a'})}{2}
~=~\frac{\sqrt{2}\,T_0}{2}
\label{newt0}
\end{equation}
where we have used the asymptotic behavior of 
$\Phi$ for $R\to\infty$
(see eq.(\ref{limi56}) and the explicit expression of the
tensions $T_p$ (see \eq{bounda3}). 
Thus, in the decompactification
limit our system is described by
\begin{eqnarray}
\ket{\widetilde B,+} &=&
\frac{\sqrt{2}\,T_0}{2}
\,\exp\biggl[-\!\sum_{n=1}^\infty \frac{1}{n}\,
\a_{-n}\cdot {S}^{(0)}\cdot \tilde \a_{-n}\biggr]
\exp\biggl[+\,\ii \!\sum_{r=1/2}^\infty
\psi_{-r}\cdot {S}^{(0)}\cdot\tilde \psi_{-r}\biggr]
\nonumber \\
&&\delta^{(9)}(q^i)\,\prod_{\mu=0}^9\ket{k^\mu=0}~~.
\label{bound0}
\end{eqnarray}
By performing the usual GSO projection, we then
obtain the complete boundary state
\begin{equation}
\ket{\widetilde B} \equiv P_{\rm GSO}\,
\ket{\widetilde B,+}= \frac{1}{2}\Big[\ket{\widetilde
B,+} -\ket{\widetilde B,-}\Big]
\label{btilde}
\end{equation}
which describes a D0-brane
in the Type IIB theory. Since there is no R-R sector, 
this D-particle is 
non-supersymmetric and non-BPS. Moreover, from eq.(\ref{bound0})
we explicitly see that
its tension (or mass) 
is a factor of $\sqrt{2}$ bigger than
the tension of the ordinary supersymmetric D-particle
of the Type IIA theory.

It turns out that the non-supersymmetric D particle is still an unstable 
configuration of type IIB theory. This is due to the fact that the absence of
a R-R sector implies the absence of GSO projection in the open string channel.
Therefore one gets an open string tachyon that denotes instability. However,
this tachyon is eliminated by the $\Omega$ projection by going from type IIB
to type I theory. In conclusion the $0$ brane that we have constructed above
is a stable configuration of type I theory.

\vspace{.5cm}
{\bf Acknowledgements} 
\vspace{.5cm}

We would like to thank M. Frau, T. Harmark, A. Lerda, R. Marotta, I. Pesando 
and R. Russo for many discussions on the subject of these lectures.

\vskip 1.3cm
\appendix{\Large {\bf{Appendix }}}
\label{appe}
\vskip 0.5cm
\renewcommand{\theequation}{A.\arabic{equation}}
\setcounter{equation}{0}
\noindent
Let $\gamma^i$ be the eight $ 16 \times 16 $ $\gamma$-matrices of
$SO(8)$. Starting from these matrices we can construct a chiral
representation for the $ 32 \times 32 $ $\Gamma$-matrices of
$SO(1,9)$, {\it i.e.}
\bea
\Gamma^{i} &=&
\pmatrix{
0 & \gamma^i \cr
\gamma^i & 0 \cr
} = \sigma^1 \otimes
\gamma^i  ~~,
\nonumber \\
\Gamma^{9} &=&
\pmatrix{
0 & \gamma^1\cdots\gamma^8 \cr
\gamma^1\cdots\gamma^8 & 0 \cr
} = \sigma^1 \otimes \left(\gamma^1\cdots\gamma^8 \right)~~,
\label{gamma} \\
\Gamma^{0} &=&
\pmatrix{
0 & \one \cr
-\one & 0 \cr
} = {\rm i} \,\sigma^2 \otimes \one   ~~,
\nonumber
\ena
where $\sigma^a$'s are the standard Pauli matrices.
One can easily verify that these matrices satisfy
$\{\Gamma^\mu\,,\, \Gamma^\nu\}=2\eta^{\mu\nu}$.
Other useful matrices are
\bea
\Gamma_{11} &=& \Gamma^0\ldots \Gamma^9=\pmatrix{
\one & 0 \cr
0 & -\one \cr
} = \sigma^3 \otimes \one  ~~,
\label{CC} \\
C &=& \pmatrix{
0 & -{\rm i}\one \cr
{\rm i}\one & 0 \cr
} = \sigma^2 \otimes \one ~~,
\nonumber
\ena
where $C$ is the charge conjugation matrix such that
\beq
\left(\Gamma^\mu\right)^T = - C\,\Gamma^\mu\,C^{-1}  ~~.
\label{transp}
\eeq
Let $A,B,...$ be 32-dimensional
indices for spinors in ten dimensions, and
$|A\rangle |{\widetilde B}\rangle$ denote the vacuum of
the Ramond fields $\psi^\mu(z)$ and
${\widetilde \psi}^\mu({\bar z})$ with spinor indices $A$ and
$B$ in the left and right sectors respectively, that is
\beq
|A\rangle |{\widetilde B}\rangle =
\lim_{z,{\bar z}\to 0} S^A(z)\,{\widetilde S}^B({\bar z})
|0\rangle
\label{vacapp}
\eeq
where $S^A$ (${\widetilde S}^B$) are the left (right)
spin fields~\cite{FMS}, and $|0\rangle$ the Fock vacuum of the
Ramond fields.
The action of the Ramond oscillators $\psi_n^\mu$ and ${\widetilde
\psi}_n^\mu$ on the state $|A\rangle |{\widetilde B}\rangle$ is given by
\beq
\psi_n^\mu \,|A\rangle |{\widetilde B}\rangle
={\widetilde \psi}_n^\mu\,
|A\rangle |{\widetilde B}\rangle
= 0
\label{psin}
\eeq
where $n$ is a positive integer and 
\bea
\psi_0^\mu\, |A\rangle |{\widetilde B}\rangle
&=& \frac{1}{\sqrt{2}} \left(\Gamma^\mu\right)^A_{~C}
\,\left(\!\one\, \right)^B_{~D}|C\rangle\, |{\widetilde D}\rangle
\nl
{\widetilde \psi}_0^\mu \,|A\rangle |{\widetilde B}\rangle
&=& \frac{1}{\sqrt{2}} \left(\Gamma_{11}\right)^A_{~C}
\,\left(\Gamma^\mu\right)^B_{~D}\,
|C\rangle |{\widetilde D}\rangle~~~.
\label{psi0}
\ena
It is easy to check that this action correctly
reproduces the anticommutation properties of the fermionic oscillators,
and in particular that
$\{\psi_0^\mu\,,\,\psi_0^\nu\}=\{{\widetilde \psi}_0^\mu
\,,\,{\widetilde\psi}_0^\nu\}=\eta^{\mu\nu}$, and
$\{\psi^\mu\,,\,{\widetilde \psi}^\nu\}=0.$
On the conjugated state $\langle A|\langle \widetilde B|$ we have analogously
\beq
\langle A|\langle\widetilde B|\,\psi_n^\mu \
=\langle A|\langle\widetilde B|{\widetilde \psi}_n^\mu\,= 0
\label{bpsin}
\eeq
if $n<0$, and
\bea
\langle A|\langle\widetilde B|\,\psi_0^\mu\,
&=& -\frac{1}{\sqrt{2}}\langle C|\langle\widetilde D|\psi_0^\mu\,
\left(\Gamma^\mu\right)^A_{~C}
\,\left(\!\one\, \right)^B_{~D}
\nl
\langle A|\langle\widetilde B|\,\psi_0^\mu\,
{\widetilde \psi}_0^\mu \,
&=& \frac{1}{\sqrt{2}}\langle C|\langle\widetilde D|
\left(\Gamma_{11}\right)^A_{~C}
\,\left(\Gamma^\mu\right)^B_{~D}\,\,\, .
\label{bpsi0}
\ena
We now use these definitions to derive the fermionic structure of the
boundary state $|B,\eta\rangle ^{(0)}_R$ in eq.(\ref{solover}), which has to 
satisfy the following overlap equation (see eq.(7.230) of Ref.~\cite{island})
\beq
\left( \psi_0^\mu -{\rm i}\eta \,{S}^\mu~_\nu {\widetilde \psi}_{0}^\nu
\right)\,|B,\eta\rangle ^{(0)}_R= 0
\label{overlap23}
\eeq
where ${S}^\mu~_\nu$ is the matrix defined in \eq{matS}.
If we write
\beq
|B_{\psi}, \eta \rangle^{(0)} = {\cal M}_{AB}\,|A\rangle |{\widetilde B}\rangle
\label{appB0}
\eeq
then, \eq{overlap23} for $n=0$ implies that the $32\times32$ matrix
${\cal M}$ has to satisfy the following equation
\beq
\left(\Gamma^\mu\right)^T\,{\cal M} - {\rm i}\,\eta
{S}^\mu\,_\nu\Gamma_{11}\,{\cal M}\,\Gamma^\nu = 0~~~.
\label{eqM}
\eeq
Using our previous definitions, one finds that a solution is
\beq
{\cal M} = C\,\Gamma^0\cdots\Gamma^p\,\frac{1+{\rm i}\eta
\,\Gamma_{11}}{1+{\rm i}\eta}~~~.
\label{defM}
\eeq
In the same way we can determine the conjugated state $\langle
B_\psi,\eta|^{(0)}_{\rm R},$ which satisfies the conjugated equation
\beq
\label{conj4b}
\langle B_\psi,\eta|^{(0)}_{\rm R}\left(\psi^\mu_{0}+i \eta S^{\mu}~_{\nu}
{\widetilde\psi^\nu_{0}}
\right)=0~~~.
\eeq
Writing 
\beq
\label{conj7b}
\langle B_\psi,\eta|^{(0)}_{\rm R}=\langle A|\langle\widetilde B| ~{\cal
N}_{AB}
\eeq
and using eqs.(\ref{bpsi0}), we can rewrite eq.(\ref{conj4b}) as
\beq
\label{nab}
-\left(\Gamma^\mu\right)^T {\cal N}+{\rm i}\,\eta S^\mu~_\nu
\left(\Gamma^{11}\right)^T{\cal N}\Gamma^\nu=0
\eeq
which is satisfied by
\beq
\label{bconj8}
 {\cal
N}_{AB}=(-1)^p\left(C\Gamma^0...\Gamma^p\frac{1+i\eta\Gamma^{11}}{1-i\eta}
\right)_{AB}
\eeq

\end{document}